\documentclass[traditabstract]{an}

\usepackage{natbib}
\usepackage[pdftex]{graphicx}
\usepackage{url}

\usepackage{caption}
\usepackage{rotating}
\usepackage{lscape}

\def\aj{AJ}%
\def\araa{ARA\&A}%
\def\apj{ApJ}%
\def\apjl{ApJ}%
\def\apjs{ApJS}%
\def\aap{A\&A}%
\def\aapr{A\&A~Rev.}%
\def\mnras{MNRAS}%
\def\pasp{PASP}%
\def\nat{Nature}%
\def\s4g{S$^4$G}

\def\hi{H{\sc I}}

\def\hi{H\,{\sc i}}

\def\h2{$\rm H_{2}$}
\def\xco{$\rm X_{CO}$}

\begin{document}
\Pagespan{1}{13}% Document's page range. 
% If second parameter is left empty, the last page is computed automatically.
\Yearpublication{----}%
\Yearsubmission{2012}%
\Month{3}%   
\Volume{---}%  
\Issue{--}% 

\title{The Opacity of Spiral Galaxy Disks IX;}
\subtitle{Dust and Gas Surface Densities}

\author{B. W. Holwerda\inst{1}\thanks{E-mail: benne.holwerda@esa.int}, R. J. Allen\inst{2}, W. J. G. de Blok\inst{3}, A. Bouchard\inst{4}, 
R. A. Gonz\'alez-L\'opezlira\inst{5}, P. C. van der Kruit\inst{6}, and A. Leroy\inst{7}
%  \thanks{\emph{Present address:}
%    European Space Agency Research Fellow (ESTEC), Keplerlaan 1, 2200 AG Noordwijk, The Netherlands}
    }

% \offprints{B.W. Holwerda, \email{benne.holwerda@esa.int}}
\titlerunning{Dust and Gas Surface Densities in Spiral Disks.}
\authorrunning{Holwerda et al.}

\institute{European Space Agency Research Fellow (ESTEC), Keplerlaan 1, 2200 AG Noordwijk, The Netherlands \and
Space Telescope Science Institute, 3700 San Martin Drive, Baltimore, MD 21218, USA \and
Stichting ASTRON, PO Box 2, 7990 AA Dwingeloo, The Netherlands
\and 
Department of Physics, Rutherford Physics Building, McGill University, 3600 University Street, Montreal, Quebec, H3A 2T8, Canada\and
Centro de Radiastronom$\acute{i}$a y Astrof$\acute{i}$sica, Universidad Nacional Aut\'onoma de M\'exico, 58190 Morelia, Michoac\'an, Mexico\and
Kapteyn Astronomical Institute, University of Groningen, PO Box 800, 9700 AV Groningen, The Netherlands\and
National Radio Astronomical Observatory, 520 Edgemont Road, Charlottesville, VA 22903, USA.}

\received{9 March 2012}
\accepted{-----}
\publonline{later}

% \abstract {context} {aims} {method} {results} {conclusions}
\abstract{%
Our aim is to explore the relation between gas, atomic and molecular, and dust in spiral galaxies. Gas surface densities are from atomic hydrogen and CO line emission maps.
To estimate the dust content, we use the disk opacity as inferred from the number of distant galaxies identified in twelve HST/WFPC2 fields of ten nearby spiral galaxies.
The observed number of distant galaxies is calibrated for source confusion and crowding with artificial galaxy counts and here we verify our results with sub-mm surface brightnesses from archival {\em Herschel-SPIRE} data.
% We aim to determine the spatial relation between the atomic, molecular, and dusty components of the ISM as traced, respectively, by \hi\ and CO column density maps and by dust extinction through the spiral disk. We compare the number of distant galaxies seen through ten nearby foreground spiral disks in twelve HST/WFPC2 images, to the \hi\ and molecular column densities from 21 cm radio observations and CO(J=2-1) maps. The observed number of distant galaxies is calibrated for source confusion and crowding with artificial galaxy counts. We find a good agreement between the local {\em Herschel} surface brightnesses and the opacity measured from the number of distant galaxies.
%
% HI comparison
We find that the opacity of the spiral disk does not correlate well with the surface density of atomic (\hi) or molecular hydrogen (${H_2}$) alone implying that dust is not only associated with the molecular clouds but also the diffuse atomic disk in these galaxies. 
% Dust-to-gas ratio
Our result is a typical dust-to-gas ratio of 0.04, with some evidence that this ratio declines with galactocentric radius, consistent with recent {\em Herschel} results.
We discuss the possible causes of this high dust-to-gas ratio; an over-estimate of the dust surface-density, an under-estimate of the molecular hydrogen density from CO maps or a combination of both.
We note that while our value of the mean dust-to-gas ratio is high, it is consistent with the metallicity at the measured radii if one assumes the Pilyugin \& Thuan calibration of gas metallicity.
}

% conclusion 1
%We find that, however we compare the number of distant galaxies --per field or per column density--, there is no correlation between the \hi\ column density and the number of distant galaxies seen through the disk. Similarly, there is no good correlation with the molecular column density, inferred from the CO maps. The lack of a relation between \hi\ column density and obscuration of distant objects implies that giant molecular clouds and \hi\ do no coincide spatially in the disk, on scales smaller than our WFPC2 fields. This does suggest that the common practice to use \hi\ maps to generate  a reddening estimate of stellar populations will result in underestimates of the reddening and dimming of stellar light in external galaxies.
% conclusion 2
% However, there is a clean radial dependence of the ratio between disk opacity -- or implied dust surface density-- and total hydrogen (\hi+\h2) surface density for our sample. The typical observed dust-to-total-gas ratio is higher than previous measurements (D2G$\sim 0.05$). This is confirmed with a comparison of our dust-to-total-gas ratio estimates to metallicity measurements from the literature.The high dust-to-gas values may be attributed to either the mean opacity overestimating the dust surface density or the CO maps underestimating the total \h2\  content of spiral disks, and most likely a combination of both.}

\keywords{Opacity, ISM: dust, extinction, ISM: structure, Galaxies: ISM, Galaxies: spiral, Galaxies: structure}

\maketitle

%\cite{Mac-Low10}Midplane Pressure and the Abundance of Molecular Hydrogen in Galaxies: Non-Equilibrium Chemical Models
%\cite{Paradis11} -Detection and characterization of a 500 mic dust emissivity excess in the Galactic Plane using Herschel/Hi-GAL observations

%\cite{Feldmann11}: on kpc scales the X-factor depends only weakly on radiation field and column density, but is still a strong function of metallicity.

\section{Introduction}

The radio 21-cm emission of atomic hydrogen (\hi) observed in the disks of spiral galaxies is a powerful tracer of the presence and dynamics of the interstellar medium (ISM), extending to well outside the typical scale of the stellar disk. Its origin is likely a mix of ``primordial" \citep{Fall80}, or recently accreted material \citep{Sancisi08}, recycled matter \citep[ejecta raining back onto the disk; e.g.,][]{Oosterloo07}, and skins of photo-dissociated material surrounding molecular clouds \citep{Allen04}. The other components of the ISM, ionised and molecular hydrogen, metals and dust, are all more difficult to trace, because their emission strengths depend on the local degree of excitation which in turn is affected by particle densities and temperatures, photon densities, and stellar and AGN illumination.

Molecular hydrogen is usually traced with CO(J=1-0 or 2-1) line emission, and from it we have derived our knowledge of the molecular clouds in nearby spirals \citep[e.g,][]{Rosolowsky05a,Leroy08}. However, it remains an open question how sensitive the CO brightness is to the local volume density and temperature of the ISM, and what is the accuracy with which observations of CO surface brightness can be converted into \h2\  column densities and ultimately into molecular cloud masses. This conversion is also likely to depend on metallicity and hence galactocentric radius \citep{Madden97, Israel97, Leroy07, Pohlen10, Leroy11, Foyle12}. 

Nevertheless, a successful and extensive description of the atomic and molecular ISM in spirals and their relation to the star-formation rate is currently being developed, using a multi-wavelength approach to estimate the star-formation rate, and high-resolution \hi\ and CO observations to characterize the ISM in individual galaxies \citep{Calzetti05, Kennicutt07, Thilker07a, Bendo10b, Foyle12}, in detail in small samples of galaxies \citep{Cortese06a, Boissier07, Bigiel08, Leroy08, Schruba11}, or in a generalized way over a population of galaxies \citep[e.g.,][]{Kennicutt98, Buat02, Bell03c, Kannappan04, West10a, Catinella10, Fabello11a}. Star-formation occurs when the combined ISM exceeds a threshold surface density \citep[although the exact threshold is still debated, see e.g.,][]{Bigiel08,Pflamm-Altenburg08}. The ratio between molecular and neutral ISM is set by the hydrostatic pressure \citep{Bigiel08, Leroy08, Obreschkow09c}.  Also, observational models of the role of photo-dissociation in the balance between atomic and molecular hydrogen have made steady progress \citep{Allen97, Allen04, Smith00,Heiner08a,Heiner08b, Heiner09, Heiner10}.

As an alternative to CO, one could use interstellar dust as a tracer of the molecular component in spiral galaxies, since it is  linked mechanically to the molecular phase \citep{Allen86,Weingartner01b},
 by mutual shielding from photo-dissociation, and the formation of molecular hydrogen on the surface of dustgrains \citep[e.g.,][]{Cazaux04b}. Interstellar dust can be traced by its emission or its extinction of starlight. 

% Dust from its EMISSION

Surface densities of dust in spirals have been obtained from spectral energy distribution models of multi-wavelength data \citep[e.g.,][]{Popescu00,Popescu02, Draine07, Boselli10}, from simple (modified) blackbody fits of far-infrared and sub-mm data \citep{Bendo08,Bendo10b,Gordon08,Gordon10} or FUV/FIR ratios \citep{Boissier04,Boissier05, Boissier07, Munoz-Mateos11}.
The aim is to estimate the typical temperature, mass, composition and emissivity of the dust, and the implied gas-to-dust ratio \citep{Boissier04, Boselli10,Munoz-Mateos09a, Munoz-Mateos09b,Pohlen10,Smith10b,Roman-Duval10, Magrini11b, Galliano11, Foyle12, Galametz12}.

The most recent {\em Herschel} results include a resolved temperature gradient in the disks of spirals \citep{Bendo10b,Smith10b, Engelbracht10,Pohlen10, Foyle12}, linked to increased illumination of the grains, notably in the spiral arms \citep{Bendo10b} and bulge \citep{Engelbracht10}. With sufficient spatial sampling, one can extract the ISM power spectrum but this is only possible with {\em Herschel} for local group galaxies \citep{Combes12}.
Based on {\em Herschel} data of the Virgo cluster,  \citep{Smith10b}, \cite{Cortese10} and \cite{Magrini11b} show the spatial coincidence and efficiency of stripping the dust  together with the \hi\ from the disks of spirals in a cluster environment.

% COMPARISON TO CO
In the comparison between the {\em Herschel} cold grain emission, and \hi\ and CO observations, the mass-opacity coefficient of dust grains appears to be too low in M33 \citep[the inner disk,][]{Braine10}, and M99 and M100 \citep{Eales10}. This is either because (1) its value is not well understood, (2) the conversion factor between CO and molecular hydrogen, \xco, is different in M99 and M100, or (3) the emissivity ($\beta$) is different at sub-mm wavelengths. 
\cite{Roman-Duval10} compare CO, \hi\ and dust in the Large Magellanic Cloud (LMC), and argue that the cause of the discrepancy cannot be a different emissivity, nor a different gas-to-dust ratio, but that CO clouds have \h2\  envelopes, hence \xco\ changes with different density environments \citep[an explanation also favored by][]{Wolfire10}. Other recent results seem to back variations in \xco; \cite{Leroy11} find a link between \xco\ and metallicity based on SED models of a few local group galaxies and the HERACLES CO survey. 
A solid result from the first {\em Herschel} observations is that the gas-to-dust ratio increases with galactocentric radius \citep{Pohlen10, Smith10b} as do \citep[][based on Spitzer data alone]{Munoz-Mateos09b, Bendo10a}.  \cite{Magrini11b} find a much lower than Galactic CO-to-\h2\ conversion factor based on the relation between metallicity and gas-to-dust ratio radial profiles of several Virgo cluster spirals.

Even with the excellent wavelength coverage of {\em Herschel}, the SED fit results remain degenerate between dust mass, temperature and emissivity \citep[see the reviews in][]{Calzetti01,Draine03}.  It is still especially difficult to distinguish between a mass of very cold (poorly illuminated) dust from dust with much different emissivity characteristics (the emissivity efficiency depends on wavelength as $\lambda^{-\beta}$ in the sub-mm regime with $\beta \ne 2$, which may be typical for very large grains).

While large masses of extremely cold dust can be ruled out with increasing confidence, the level of illumination of the grains by the interstellar radiation field remains a fully free parameter in the SED models. The main uncertainty is complex relative geometry between the dusty filamentary structures and the illuminating stars. Both the grain emissivity and dust/star geometry can be expected to change significantly throughout the disk, i.e., with galactocentric radius or in a spiral arm.

% OR DUST ABSORPTION
Alternatively to models of dust emission, one can use the absorption of stellar light to trace dust densities. The advantages are higher spatial resolution of optical wavelengths and an independence of dust temperature. However, one needs a known background source of stellar light to measure the transparency of a spiral disk\footnote{We used the term ``opacity" throughout our project and its publications for historical reasons.}.
Two observational techniques have been developed to measure the opacity of spirals and consequently their dust content. The first one uses occulting galaxy pairs \citep[][Holwerda et al. {\it submitted.}]{Andredakis92, Berlind97, kw99a, kw00a, kw00b, kw01a, kw01b, Elmegreen01, Holwerda07c, Holwerda09, Keel11}, of which an increasing number are now known thanks to the Sloan Digital Sky Survey and the GalaxyZOO citizen science project \citep{Lintott08}. 

The second method uses the number of distant galaxies seen through the disk of a nearby face-on spiral, preferably in Hubble Space Telescope ({\em HST}) images. The latter technique is the focus of our ``Opacity of Spiral Galaxies'' series of papers \citep{Gonzalez98, Gonzalez03, Holwerda05a, Holwerda05b, Holwerda05c, Holwerda05e, Holwerda05d, Holwerda07a}.\footnote{Other authors have used distant galaxy counts or colours to estimate extinction in the Magellanic Clouds \citep{Shapley51, Wesselink61b, Hodge74, Hodge75, McGillivray75, Gurwell90, Dutra01} and other galaxies \citep{Zaritsky94,Cuillandre01}.} The benefit of using distant galaxies as the background light source is their ubiquity in HST images of nearby galaxies.
Now that uniform \hi\ maps are available from the THINGS project \citep[The \hi\ Nearby Galaxy Survey,][]{Walter08}, as well as public {\em Herschel} data from the KINGFISH \citep[Key Insights on Nearby Galaxies: a Far-Infrared Survey with Herschel,][]{Skibba11, Dale12, Kennicutt11,Galametz12}, and CO(J=2-1) maps from the HERACLES survey \citep[The HERA CO Line Extragalactic Survey,][]{heracles} for a sub-sample of the galaxies analysed in our ``Opacity of Spiral Galaxies'' project, we are taking the opportunity to compare our disk opacities to \hi\ and \h2 surface densities to see how they relate.

Our method of determining dust surface densities is certainly not without its own uncertainties (notably cosmic variance, see \S \ref{s:sfm}) 
but these are not the ones of sub-mm emission suffers from (grain emissivity, level of stellar illumination, variance within the disk or these). 
Hence, our motivation for our comparison between the disk opacity and the other tracers of the cold ISM is to serve as an independent check to the new {\em Herschel} results.

% To trace the cold, dusty molecular component of spiral disks is our main motivation but, for this paper, we have a secondary science driver. It is not uncommon that \hi\ maps are used, together with the canonical Galactic relation between extinction and \hi\ column density reported in \cite{Bohlin78}, to produce a reddening map for an external galaxy \cite[e.g.][]{Ibata09,Whaley09}. Apart from the problem of the relative geometry of the \hi\ and the stellar population, we suspect that the \hi\ may not trace the dusty component very well. Our second motivation is to check the applicability of an \hi\ map as a reddening estimate.

In section 2, we discuss the origin of our sample and data. Section 3 explains how we derive a disk opacity from the number of distant galaxies. In section 4, we discuss the distant galaxy number as a function of \hi\ column density and in section 5, we compare the \hi\ and \h2\ column densities, dust extinction, averaged over whole WFPC2 fields, and per \hi\ contour, respectively. Sections 6 and 7 contain our discussion and conclusions.

%__________________________________________________________________

\section{Galaxy Sample and Data}

Our present sample is the overlap between the \cite{Holwerda05b},  the THINGS \citep{Walter08}, and the HERACLES \citep[][]{heracles} projects. 
The common 10 disk galaxies are listed in Table \ref{t:info}. We use the public THINGS data and early science release data from HERACLES.

%  zWe use two uniform sets of observations for this sample: 21-cm \hi\ emission maps from the VLA, and HST/WFPC2 I-band images for disk opacity estimates. 
Figure \ref{f:himap} shows the HST/WFPC2 ``footprints'' overlaid on the VLA HI maps. 
In the case of NGC 3621 and NGC 5194, there are two HST/WFPC2 fields available for each galaxy.

\subsection{VLA 21-cm Line Observations}
\label{s:hi}

For this study we use the THINGS \citep[The \hi\ Nearby Galaxy Survey,][]{Walter08} robustly-weighted (RO) integrated total \hi\ intensity maps (available from \url{http://www.mpia-hd.mpg.de/THINGS/}). The maps were obtained with the VLA, and converted to \hi\ surface density using the prescription from \cite{Walter08}, equations 1 and 5, and Table 3. Although the naturally-weighted maps are markedly more sensitive to the largest scale \hi\  distribution, the robust maps have the highest angular resolution. 

The robust maps are better suited for a direct comparison with the number of background galaxies, as we are interested in the \hi\ column density at the position of each background galaxy and hence at scales smaller than the FOV of the HST/WFPC2 FOV (3 CCDs of $1\farcm3 \times 1\farcm3$). Additionally, we use the WFPC2
footprint as an aperture on the \hi\ maps (Figure \ref{f:himap}).\footnote{In this case it does not matter whether the maps are robustly weighted or naturally weighted.} The \hi\ column densities averaged over the WFPC2 footprints (an angular scale of $2\farcm3$) on the sample galaxies, and expressed in units of ${\rm M_\odot/pc^2}$ are listed in Table \ref{t:info}. These mean column densities include a correction factor (1.36) for Helium contribution to the atomic gas phase.

 \begin{landscape}
% FIGURE 1
  \begin{figure}
   \centering
      \includegraphics[width=\textwidth]{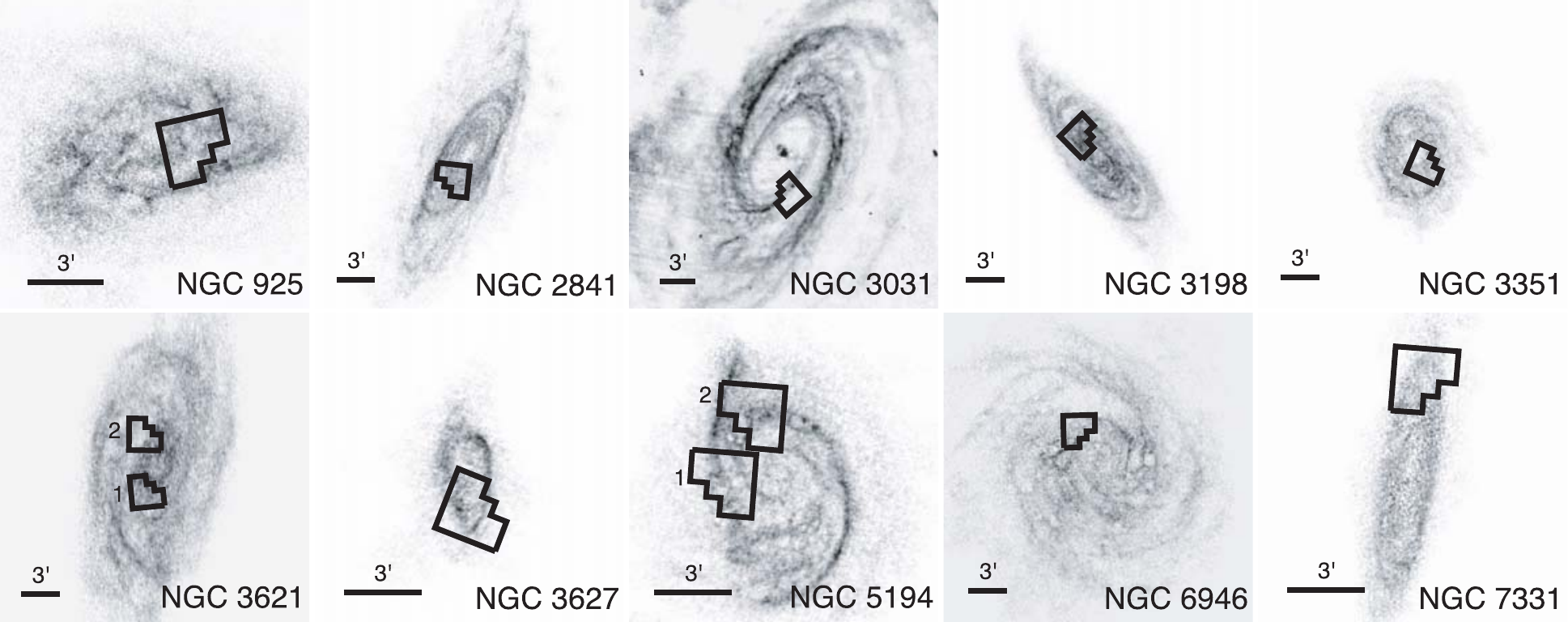}
   \caption{The THINGS robustly weighted integrated \hi\ column density maps. The HST/WFPC2 footprint is overlaid (black outline). NGC 3621 and NGC 5194 have two HST pointings each. A 3 arcminute ruler is shown for scale comparison. Most of the WFPC2 fields in \protect\cite{Holwerda05} were originally taken for the Cepheid Distance Scale Key Project (\protect\cite{KeyProject}); they were positioned on spiral arms in the outer, less crowded, parts of the disks to aid in the identification of Cepheid variables. NGC 3031 and NGC 3621 do not have CO observations.}%
   \label{f:himap} 
    \end{figure}
% \newpage
 \begin{table}
%\begin{sidewaystable*}
 {\tiny
\captionof{table}{Some basic data, our measurements in the WFPC2 fields and implied surface densities for our sample of galaxies: name (1); distance (2) from \protect\cite{mythesis}; the 25 $B$-mag arcsec$^{-2}$ radius, $R_{25}$ (3), from \protect\cite{RC3}; galactocentric radius of each WFPC2 field (4), expressed as a fraction of $R_{25}$; \hi\ surface density (5): \hi\ data linear resolution (6); mean CO luminosity (7); implied \h2\ surface density (8); CO linear resolution (9); mean opacity (10); implied dust surface density (11); and linear size of the side of one the Wide Field chip (one of the three squares in the L-shape of the WFPC2, Figure \ref{f:himap}) for each WFPC2 field (12). Metallicity ($12+log(OH)$) estimated from the linear relation with radius presented in \protect\cite{Moustakas10} for the position of each aperture. They present metalicity values based on the calibration from \protect\cite{KK04} (13) and \protect\cite{PT05} (14). {\em Herschel-SPIRE} surface brightnesses in each WFPC2 field for 250 $\mu$m (15), 350 $\mu$m (16), and 500 $\mu$m (17). There is no CO information on NGC 3031 and NGC 3621. NGC 925 and NGC 5194-1 have negative opacities, most likely due to cosmic variance in the number of background galaxies.  }
\label{t:info}     
\centering        
\begin{tabular}{l l l l l l l l l l l l l l l l l}
\hline\hline                
Galaxy		& Dist. 	& $R_{25}$ & $R/R_{25}$ & $\rm \Sigma_{HI}$ 	& $\rm FWHM$		& $L_{CO}$	& $\rm \Sigma_{H2}$	& $\rm FWHM$			& $\rm A_{SFM}$	& $\rm \Sigma_d$		& $\rm FOV$ 		& log(OH) & + 12 				&  &&\\
			&		&		&			&					& (\hi)			&			& 					& (CO)				&				&					&					& & 				& 	&&\\
			& (Mpc) 	&  (kpc) 	& 			& ($\rm M_\odot$		& (kpc) 			& (K			& ($\rm M_\odot$		& (kpc) 				& (mag.)			& ($\rm M_\odot $		& (kpc)				& KK04 & PT05 	& & & \\				
			&		&		& 			& $\rm / pc^2$)			&				& km/s)		& $\rm / pc^2$)			& 					&				& $\rm / pc^2$)			& 					& & 				& & &\\
	(1)		& (2)	    	& (3)		& (4)			& (5) 				& (6)				& (7)			& (8) 				& (9)					& (10)			& (11)				& (12)  				& (13) & (14) 		& (15) & (16) & (17)	\\
\hline                

   NGC925 & 11.20 &  5.61 &  0.50 & 17.92 &  0.33 &  0.23 &  1.83 &  0.60 & $  -0.4^{ 0.3}_{ 0.3}$ & -0.44 &  4.22 &  8.20 &  8.70 & -3.49  & -3.50 & -3.54 \\
   NGC2841 & 14.10 &  4.07 &  0.61 &  8.98 &  0.41 &  1.00 &  7.85 &  0.75 & $   0.8^{ 0.4}_{ 0.5}$ &  0.88 &  5.32 &  8.40 &  9.20 & -3.17  & -3.27 & -3.39 \\ 
   NGC3031 &  3.60 & 13.46 &  0.25 &  4.47 &  0.10 & \dots & \dots &  0.19 & $   0.8^{ 0.6}_{ 0.5}$ &  0.88 &  1.36 &  8.50 &  9.10 & -4.05  & -4.16 & -4.22 \\ 
   NGC3198 & 13.80 &  4.26 &  0.62 & 18.15 &  0.40 &  0.31 &  2.43 &  0.74 & $   0.8^{ 0.3}_{ 0.3}$ &  0.88 &  5.20 &  8.30 &  8.80 & -3.60  & -3.63 & -3.69 \\ 
   NGC3351 & 10.00 &  3.71 &  0.51 &  3.52 &  0.29 &  1.04 &  8.19 &  0.53 & $   1.2^{ 0.5}_{ 0.6}$ &  1.32 &  3.77 &  8.60 &  9.20 & -3.05  & -3.18 & -3.33 \\
 NGC3621-1 &  6.64 &  6.15 &  0.37 & 18.80 &  0.19 & \dots & \dots &  0.35 & $   2.2^{ 0.6}_{ 0.6}$ &  2.42 &  2.50 &  8.30 &  8.90 & -3.25  & -3.34 & -3.41 \\ 
 NGC3621-2 &  6.64 &  6.15 &  0.38 & 12.72 &  0.19 & \dots & \dots &  0.35 & $   1.0^{ 0.3}_{ 0.4}$ &  1.10 &  2.50 &  8.30 &  8.90 & -3.25  & -3.54 & -3.63 \\ 
   NGC3627 & 10.05 &  4.56 &  0.52 &  9.15 &  0.29 &  3.93 & 30.98 &  0.54 & $   2.1^{ 0.7}_{ 0.7}$ &  2.31 &  3.79 & \dots & \dots & -2.67  & -2.86 & -3.01 \\ 
 NGC5194-1 &  8.40 &  5.61 &  0.62 &  8.97 &  0.24 &  3.38 & 26.66 &  0.45 & $  -0.4^{ 0.4}_{ 0.4}$ & -0.44 &  3.17 &  8.50 &  9.10 & -2.88  & -3.05 & -3.21 \\ 
 NGC5194-2 &  8.41 &  5.61 &  0.63 & 10.11 &  0.24 &  4.47 & 35.26 &  0.45 & $   1.4^{ 0.6}_{ 0.6}$ &  1.54 &  3.17 &  8.50 &  9.10 & -2.84  & -3.01 & -3.14 \\ 
   NGC6946 & 11.48 &  5.74 &  0.75 &  9.44 &  0.33 &  2.12 & 16.69 &  0.61 & $   1.1^{ 0.5}_{ 0.6}$ &  1.21 &  4.33 &  8.30 &  8.90 & -3.28  & -3.44 & -3.60 \\ 
   NGC7331 & 14.72 &  5.24 &  0.76 & 21.53 &  0.43 &  0.26 &  2.06 &  0.78 & $   0.3^{ 0.3}_{ 0.3}$ &  0.33 &  5.55 &  8.30 &  8.80 & -3.57  & -3.61 & -3.69 \\

   \hline 
\end{tabular}

}
 \end{table}
%\end{sidewaystable*}
\end{landscape}
%__________________________________________________________________

\subsection{CO(J = 2 $\rightarrow$ 1) Line Observations}
% HERACLES CO J = 2 ? 1 i
\label{s:co}

The HERACLES project \citep[The HERA CO Line Extragalactic Survey,][]{heracles, Walter09} is a project on the IRAM 30m telescope to map the molecular gas over the entire optical disks ($R_{25}$) of 40 nearby galaxies via the CO(J=2-1) emission line. The HERA instrument has comparable spatial (11") and velocity (2.6 km/s) resolutions to the THINGS survey, and good sensitivity ($3 \sigma \approx 3 \rm M_\odot / pc^2$) as well. The HERACLES sample overlaps by design with the THINGS and SINGS \citep[{\em Spitzer} Infrared Nearby Galaxy Survey,][]{SINGS} samples and it also has 8 galaxies in common with our previous work (Table \ref{t:info}).

To convert the CO (J=2-1) maps to molecular hydrogen surface density maps, we need the conversion factor \xco (alternatively denoted as $\rm \alpha_{CO}$). For the CO (J=1-0) line, this is commonly assumed to be 4.4. The ratio between the CO(J=1-0) and CO(J=2-1) line is 0.7 according to the HERACLES observations. To convert the CO(J=2-1) map (in K km/s) into molecular surface density, $\rm X_{CO (2-1)} = 4.4/0.7 = 6.3 M_\odot/pc^2$ \citep{Leroy08}. The mean values of the CO(J=2-1) surface brightness and the molecular hydrogen surface density are listed in Table \ref{t:info}.

\subsection{HST/WFPC2 Images}
\label{s:hst}

The background galaxy counts are based on HST/WFPC2 data, as presented in \cite{Holwerda05} and \cite{Holwerda05b}. The footprints of the 12 HST/WFPC2 fields on the integrated \hi\  maps of 10 THINGS galaxies are shown in Figure \ref{f:himap} and we only consider these areas of the disks. The HST fields are predominantly from the Distance Scale Key Project \citep{KeyProject}, and are therefore usually aimed at spiral arms in the outer parts of the main disks, in order to facilitate the identification of Cepheids. The final drizzled WFPC2 images in $F814W$ and $F555W$, from \cite{Holwerda05}, can be obtained at \url{http://archive.stsci.edu/prepds/sgal/} and the NASA Extragalactic Database.\footnote{Similar quality products are now also available from the archives at STSCI, the High-Level Archive; \url{www.hla.stsci.edu}.}

\section{Disk Opacity from the Number of Background Galaxies.}
\label{s:sfm}

The central premise of our method to measure disk opacity, is that the reduction in the number of distant galaxies seen though a foreground spiral galaxy is a reasonable indication of the transparency of the disk. The number of distant galaxies that can be identified is a function of several factors: the real number of galaxies behind the disk; the crowding by objects in the foreground disk and consequently the confusion in the identification of the distant galaxies, and, finally, absorption of the light from the background galaxies by the interstellar dust in the foreground disk. Since we are only interested in the last one --the dust extinction--, all the other factors need to be mitigated and accounted for. {\em HST} provides the superb resolution to identify many distant galaxies, even in the quite crowded fields of nearby spiral galaxies. But to fully calibrate for crowding and confusion, we developed
the ``Synthetic Field Method'' (SFM), in essence a series of artificial galaxy counts under the same conditions as the science field \citep{Gonzalez98, Holwerda05a}.    
   
If we identify $N$ galaxies in a field, we need to know two quantities to convert this number into a disk opacity measurement: (1) the number ($N_0$) of galaxies we would have identified in this field, without any dust extinction but under the same crowding and confusion conditions, and (2) the dependence ($C$) of the
number of galaxies on any increase of dust extinction. The disk's opacity in $F814W$ is then expressed as:
\begin{equation}
\label{eq1}
A_I = -2.5~ C~ \log \left({N \over N_0}\right).
\end{equation}

If the number of identified galaxies behaved exactly as photons, the parameter $C$ would be unity. We have found it to be close to 1.2 for a typical field, and $N_0$ to depend the surface brightness and granularity of the foreground disk \citep{Gonzalez03,Holwerda05e}. From our artificial distant galaxy counts in the WFPC2 fields, we can obtain both $N_0$ and $C$; the first from an artificial count of seeded, undimmed, distant galaxies, and the second from a series of artificial distant galaxy counts with progressive dimming of the  seeded galaxies. 

Since we cannot know the intrinsic number of distant galaxies behind the foreground disk, we treat the cosmic variance as a source of uncertainty  in $N_0$ that can be estimated from the observed 2-point correlation function. This typically is of the same order as the Poisson error in the opacity measurement.\footnote{It depends to a degree on the depth of the data. Conservatively, for this kind of fields, the total error is about 3.5 times Poisson \citep{Gonzalez03}.} Because the cosmic variance uncertainty is substantial, improvements in the identification of distant galaxies barely improve our errors \citep[see also][]{mythesis}. 

To test the general SFM results, we have done several checks against other techniques. 
% OCG
The results are consistent with those obtained from occulting galaxy pairs \citep{Holwerda05b}, both the results in \cite{kw00a,kw00b} as well as the later opacities found in \cite{Holwerda07c}.
% cepheid reddening.
The SFM results are also consistent with the amount of dust reddening observed for the Cepheids in these fields \citep[the majority of which is from the Cepheid Distance Scale Key Project ][]{KeyProject}, the dust surface densities inferred from the far-infrared SED \citep[][discussed below]{Holwerda07a}, and the sub-mm fluxes from KINGFISH observations (\S \ref{s:herschel}, below).
Even with HST, the number of identifiable galaxies in a given WFPC2 field is relatively small, a fact that results in large uncertainties if the field is further segmented for its analysis, e.g., sub-divided into arm and inter-arm regions. To combat the large uncertainties, we  combined the numbers of background galaxies found in different fields, based on certain characteristics of the foreground disks, like galactocentric radius, location in the arm or inter-arm regions \citep{Holwerda05b}, surface brightness \citep{Holwerda05d}, or NIR colour \citep{Holwerda07b}. Because no uniform \hi\ and CO maps were available until now, we compared radial \hi\ profiles to our radial opacity profile in \cite{Holwerda05c}, but this is far from ideal. Now that the THINGS and HERACLES maps are available, we can compare the average opacity of an {\em HST} field to its mean \hi\ and \h2\ surface densities or, alternatively, rank the distant galaxies based on the foreground disk's \hi\ column density at their position.

\subsection{Dust Surface Densities}
\label{s:sigmad}

To convert the above opacity of the spiral disk to a dust surface density, we assume a smooth surface density distribution of the dust (no clumps or fine structure). 
The dust surface density is then:
\begin{equation}
\Sigma_{\rm d}  = {1.086 A  \over \kappa_{\rm abs}},
\end{equation}
\noindent with $\kappa_{\rm abs}$ for Johnson $I$ from \cite{Draine03}, Table 4; $4.73 \times 10^3$ cm$^2$ g$^{-1}$. The mean opacity ($A_{\rm SFM}$) and implied mean dust surface densities are listed in Table \ref{t:info}. The value for $\kappa_{\rm abs}$ changes with the types of grain (and hence with environment in the disk) and the Draine et al. is a value typical for large grains. Variance in $\kappa_{\rm abs}$ is not unusual depending on the prevailing composition of the dust.

% effects of clumpiness
%
The screen approximation to estimate the surface density is common but in fact the dusty ISM is clumped and filamentary in nature with a wide range of densities and temperatures. Typically, the distant galaxies are seen in gaps between the dusty clouds \citep{Holwerda07b}. 
The typical value of $A_I \sim 1$ (Figure \ref{f:sighi-A}) corresponds to a surface covering factor of 60\%, if the clouds were completely opaque. In reality, the disk opacity is a mix of covering factor and the mean extinction of the clouds \citep[on average $\tau_{\rm cloud} = 0.4$ and cloud size 60 pc,][]{Holwerda07a}. 
We note that our mass estimates agree with those from a fit to the {\em Spitzer} fluxes with the \cite{Draine01b} model \citep[to within a factor of two][Figure \ref{f:draine}]{Holwerda07a}. \cite{Draine07} note that the addition of sub-mm information to such a fit may modify the dust mass estimate by a factor of 1.5 or less. Thus, while there is certainly a range of dust densities in each field, we are confident that  the estimate from the above expression is a reasonable {\em mean} surface density.

\subsection{Herschel-SPIRE Surface Brightness}
\label{s:herschel}

Sub-mm data for all our galaxies are available at the {\em Herschel} Science Archive\footnote{\url{http://herschel.esac.esa.int/Science_Archive.shtml}}, the majority taken for the KINGFISH project\footnote{Key Insights on Nearby Galaxies: a Far-Infrared Survey with Herschel, PI. R. Kennicutt, \citep[see also][]{Skibba11a, Dale12, Kennicutt11, Galametz12}}.
We therefore check the reliability of the SFM as a tracer of the dust surface density by directly comparing the surface brightness measured by the 
Spectral and Photometric Imaging REceiver \citep[SPIRE,][]{Griffin10}, onboard {\em Herschel} to the opacity as measured by the SFM.

\begin{figure}
   \centering
    \includegraphics[width=0.5\textwidth]{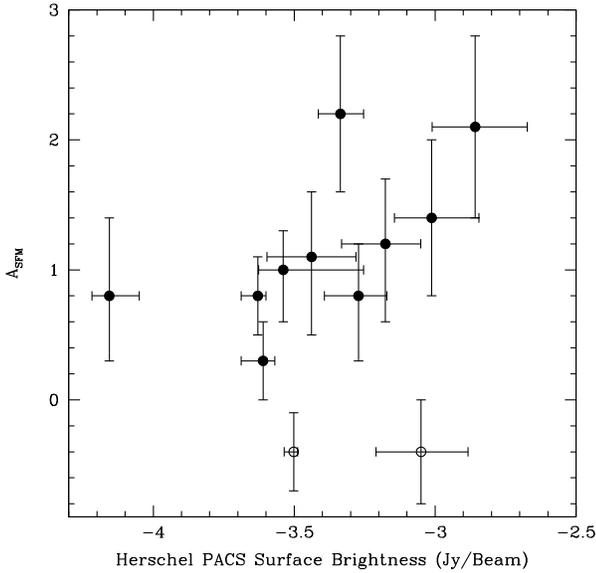}
         \caption{The Herschel/SPIRE 350 $\mu$m mean surface brightnesses in the WFPC2 field-of-view. Horizontal bars mark the 250 (left) and 500 (right) $\mu$m fluxes in the same field. 
         The width of the horizontal bar is indicative of the mean temperature of the dust in each disk (a wide bar points to higher mean temperature).
         Variance around the mean surface brightness in each band is substantial due to both Poisson noise and structure in the galaxy disk. 
         The lowest surface brightness point is NGC 3031, the closest galaxy in our sample. 
         This field is right on the edge of the ISM disk (Figure \ref{f:himap}) and therefore suffers the most from uncertainties due to internal structure and aperture correction. }
     \label{f:herschel}
\end{figure}

We used the WFPC2 field-of-view as the aperture to measure the fluxes at 250, 350, and 500 $\mu$m (listed in Table \ref{t:info}), similar to our measurements of the surface density in the \hi\  and CO data (\S \ref{s:hi} and \ref{s:co}). These were not aperture corrected because of the unique shape of the aperture. Figure \ref{f:herschel} shows the {\em Herschel} surface brightnesses versus the SFM opacities for all three wavebands (the 250 and 500 $\mu$m. values are the end points of the horizontal bars). To convert the flux in a {\em Herschel-SPIRE} waveband into a dust surface density, one would need both a typical dust temperature or a temperature distribution and the dust's emissivity. The horizontal bars indicate there is a range of mean temperatures in these disks.

\begin{figure}
   \centering
    \includegraphics[width=0.5\textwidth]{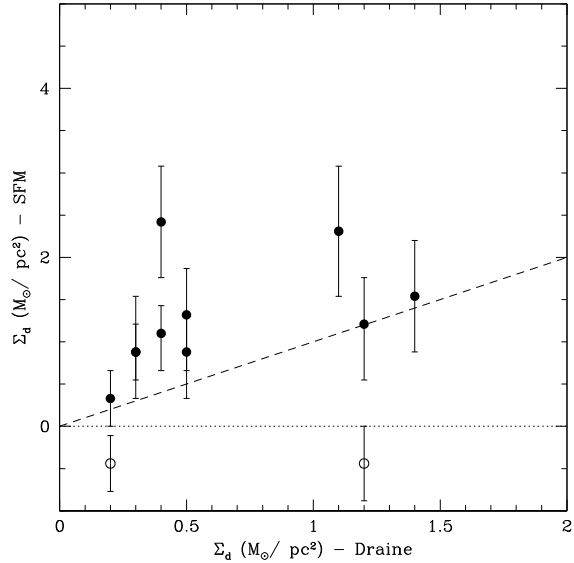}
         \caption{The dust surface density inferred by the SED model from \cite{Draine07} based on Spitzer fluxes \citep[presented earlier in][]{Holwerda07a} compared to those from the SFM. There is at most a factor two difference between these, consistent with the lack of sub-mm information in these initial fits. Dashed line is the line of equality. }
     \label{f:draine}
\end{figure}

There is a linear relation between the {\em Herschel-SPIRE} surface brightnesses and the SFM opacities. The scatter is much less for this relation than between the SFM dust surface density values and those inferred from far-infrared SED models \citep[][Figure \ref{f:draine}]{Holwerda07a}. Hence, we conclude that the SFM opacities are a reasonable indicator for mean dust surface density.

As a qualitative check, we compare the dust surface densities derived for a subset of the KINGFISH sample by \cite{Galametz12}, their Figure A1, to those derived above. Typical values mid-disk for the overlap (NGC 3351, NGC 3521 and NGC 3627), where the WFPC2 images are located, are $\sim0.3 M_\odot/pc^2$, which appear to be typical \citep[i.e., similar to those in][]{Foyle12}. These values lie a factor two below the ones implied by the SFM (Table \ref{t:info}), regardless of the dust emissivity used in the \cite{Galametz12} fits but the difference is greater for fits where the emissivity is a free parameter. We found similar dust surface densities from the SED model in \cite{Holwerda07a}, based on the {\em Spitzer} fluxes alone (Figure \ref{f:draine}). Because all these models are based on the \cite{Draine07} model, we made a second check using the {\sc magphys} SED model ({\sc magphys}). These dust surface density are to a factor ten below the SFM or Draine et al. estimates. These fits illustrate the importance of the choice of model compared to the inclusion of sub-mm data.

\section{\hi\ Column Density and the Number of Distant Galaxies}
\label{s:hi}

To improve statistics, one our tactics has been to stack the numbers of galaxies in our fields according to a local characteristic (surface brightness, galactocentric radius etc.). Here we combine the number of background galaxies, both real and artificial, based on the \hi\ column density at their respective positions. 
% One approach to establish the relation between disk opacity and \hi\ column density is to determine the \hi\ column density at the position of each identified distant galaxy\footnote{The Helium fraction correction factor was not applied here.}. This can be done for both real and artificial --without any dimming applied-- distant galaxies. The distant galaxies can now be sorted according to \hi\ column density at their position. 
If there is a relation between disk opacity and \hi\ column density resolved in the THINGS RO maps, it should show as a preference of the real distant galaxies for a specific \hi\ column density, for example for lower values of $\rm \Sigma_{HI}$. The artificial galaxies would not prefer any \hi\ column density value in particular. 
% We grouped all the counts in our sample together in order to improve statistics. 
%
% FIGURE 2
\begin{figure}
   \centering
    \includegraphics[width=0.5\textwidth]{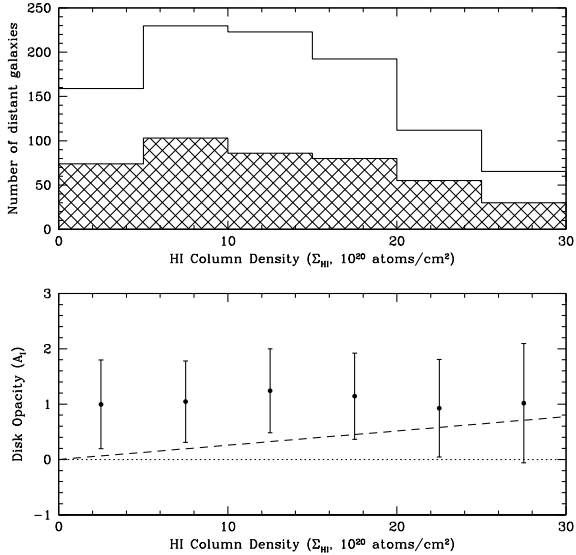}
      \caption{{\bf Top:} histogram of real (hatched) and artificial galaxies, $N$ and $N_0$ respectively, as a function of \hi\  surface density, $\rm \Sigma_{HI}$.
Because all the WFPC2 fields were chosen on spiral arms at the edge of the optical disks, the range of $\rm \Sigma_{HI}$ is limited. {\bf Bottom:} inferred opacity ($A_I$) as a function of \hi\ surface density. The dashed line is the relation from {\protect\cite{Bohlin78}} for the Galactic total (\hi+\h2) gas-to-dust ratio. }
         \label{f:na}
   \end{figure}
The top panel in Figure \ref{f:na} shows the distribution histogram of real (hatched) and artificial (solid) galaxies observed, as a function of foreground galaxy \hi\ column density. The bottom panel converts the ratio of real and artificial galaxies found at  an \hi\ column density into an opacity, using equation \ref{eq1} with C equal to 1.2. The real distant galaxies identified in the HST images do not show a clear preference for a certain \hi\ column density. Their distribution is very similar to that of the artificial distant background galaxies. As a result, the inferred opacity is constant with \hi\ column density. 
In our opinion, this lack of a relation can either be: (1) real, pointing to a break-down in the spatial relation between \hi\ and dust on scales of 6$^{\prime\prime}$ (corresponding to $\sim0.5$ kpc in our galaxies); or (2) an artifact of stacking results from different fields at various galactocentric radii in different foreground galaxies at diverse distances.
% various different galaxies which are observed at different distances and with fields at different radii. 
We note, however, that the deviation from the \cite{Bohlin78} relation between column density and extinction (dashed line in bottom panel) is strongest for the lowest \hi\ column densities, where our statistics are the most robust. In our opinion, this points to that one needs to compare to the total hydrogen column density, including the molecular component\footnote{Our fields are usually centered on a spiral arm (to observe Cepheids) and this increases the contribution from molecular phase.}.
% An \hi\ column density is not enough, one needs an estimate for the molecular hydrogen as well.

% FIGURE 3
\begin{figure*}
\centering
\includegraphics[angle=0, width=0.32\textwidth]{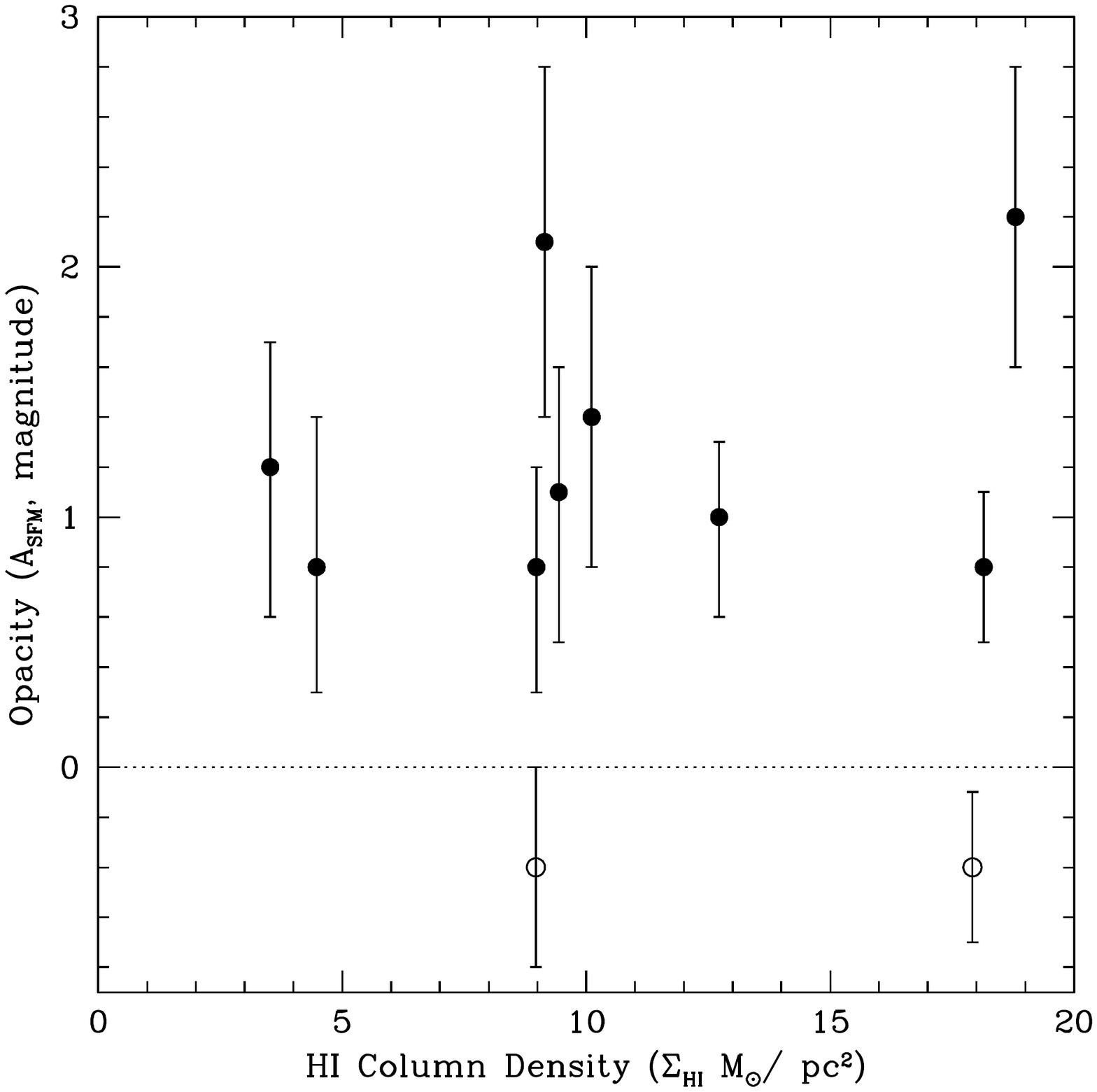}
\includegraphics[angle=0, width=0.32\textwidth]{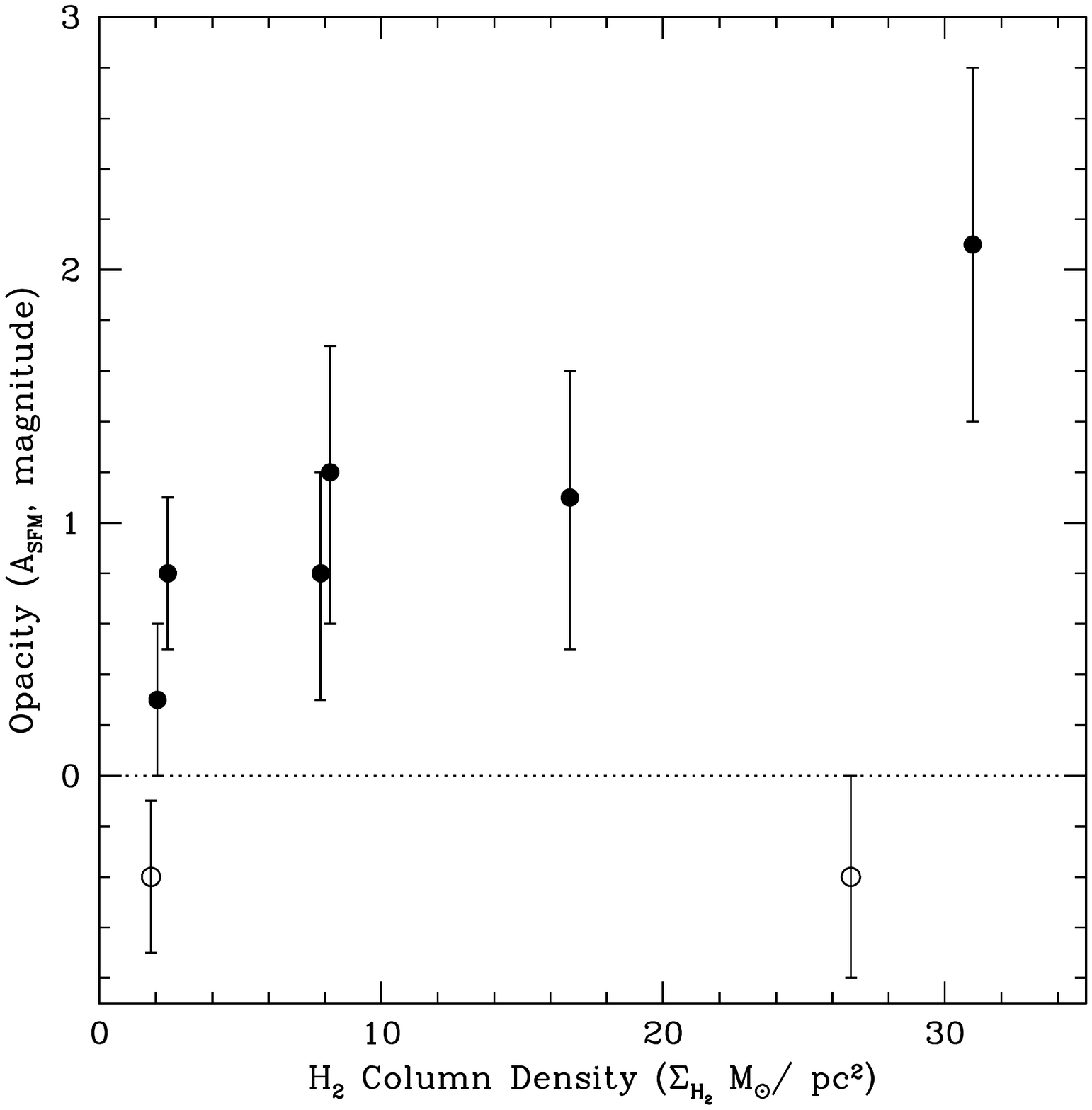}
\includegraphics[angle=0, width=0.32\textwidth]{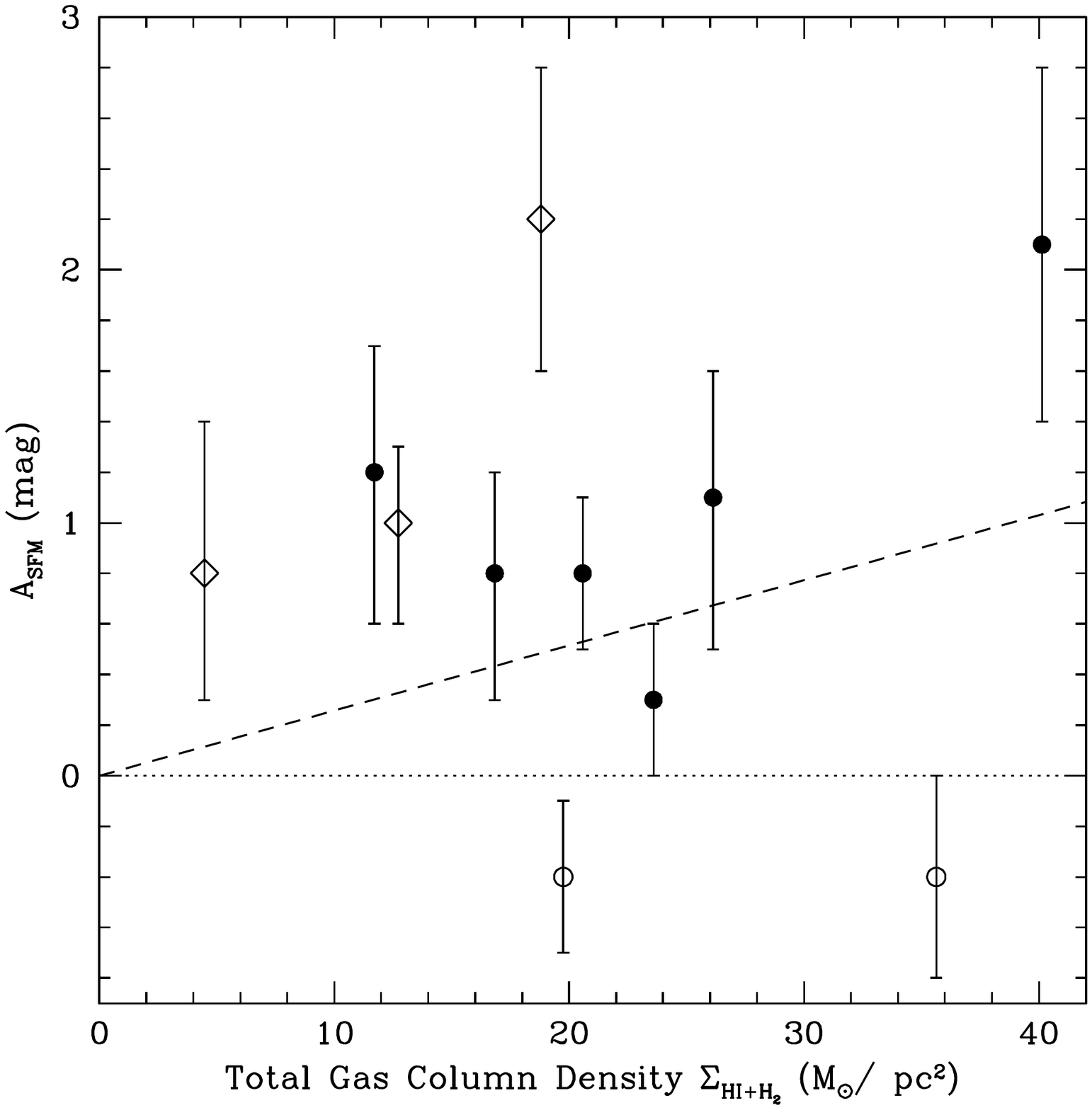}
\caption{The relation between mean \hi, \h2 and total gas (\hi+\h2) column density ($\rm \Sigma_{HI}$) and average opacity ($\rm A_{SFM}$) for each WFPC2 field. Two fields are on average negatively opaque, an effect of cosmic variance in the field of galaxies behind them (open circles) and those fields without $\rm H_2$ information denoted by open diamonds. There is no clear relation between \hi\ surface density and the opacity of the WFPC2 fields, and only a hint of a relation between \h2\ surface density and disk opacity. Most points lie above the the relation found by Bohlin et al. (1978), {\it dashed line}, right.}
\label{f:sighi-A}
\end{figure*}

\section{Average Column Densities and Opacity per WFPC2 field}
\label{s:wfpc2}

% DISK OPACITIES  VS GAS SURFACE DENSITIES
Our second approach is to compare \hi\ and \h2\ column densities to disk opacity averaged over each WFPC2 field. Table \ref{t:info} lists the average opacity value for each HST/WFPC2 field, and the \hi\ and \h2\ column densities averaged over the WFPC2 field-of-view (the footprints in Figure \ref{f:himap}). The beams of the \hi\ and \h2\ observations are much smaller than the WFPC2 apertures and we expect any aperture correction to the surface densities to be small (Table \ref{t:info})\footnote{We chose not to correct the surface densities because of the odd shape of the aperture. Depending on how one treats the edges of the aperture, the average surface density varies with $\sim$10\%.}.
% HI
Figure \ref{f:sighi-A}, left, plots the opacity versus \hi\ surface density; there is no clear relation between the two, when averaged over the size of a WFPC2 field. There are two negative values in our present sample [and the entire \cite{Holwerda05} sample],  that are probably due to cosmic variance in the number of background galaxies (a background cluster). The opacity values and \hi\ surface densities span a reasonable range for spiral galaxies. \cite{Cuillandre01} similarly find little relation between reddening and number of distant galaxies, on one side, and \hi\ column density, on the other.
% H2
Figure \ref{f:sighi-A}, middle panel, shows the relation between disk opacity and mean surface density of \h2, inferred from the CO observations. There are fewer useful points, as there are no CO data for three of our WFPC2 fields, and two of the WFPC2 fields show the aforementioned negative opacity. There could be a relation between CO inferred molecular surface density and opacity. 
% HI+H2
Figure \ref{f:sighi-A}, right panel, shows the relation between disk opacity and mean surface density of total gas (\hi+\h2). For those galaxies where no CO information as available, we use the \hi\ mean surface density (open diamonds). For comparison, we show the canonical Galactic relation from Bohlin et al. Opacity appears mostly independent from total gas surface density but with the majority of our points lie above the Bohlin et al. relation. 
There is surprisingly little of a relation between the gas, total, molecular or atomic, and disk opacity. 
In part this may be due in part to the different dust clumpiness in each disk, which is observed at a different distance. 
Alternatively, the metallicity and implicitly the average galactocentric radius of each field is the missing factor in the gas-dust relation in these fields.

% DISK OPACITIES  VS GAS SURFACE DENSITIES AS A FUNCTION OF RADIUS

%
%%% FIGURE 4
%\begin{figure}
%\centering
%\includegraphics[angle=0, width=0.5\textwidth]{./holwerda_f5.pdf}
%\caption{ The ratio between average opacity ($\rm A_{SFM}$) and mean \hi\ column density ($\rm \Sigma_{HI}$) of each WFPC2 field, as a function of radial position in the disk. Two fields are on average negatively opaque, probably as a consequence of cosmic variance in the number of galaxies behind them. }
%\label{f:sigA}
%\end{figure}
%
%
%% FIGURE 5
%\begin{figure}
%\centering
%\includegraphics[angle=0, width=0.5\textwidth]{./holwerda_f6.pdf}
%\caption{Surface densities (dust $\rm \Sigma_{d}$, \h2\ $\rm \Sigma_{H_2}$ and \hi\ $\rm \Sigma_{HI}$) as a function of galactocentric radius, in units of the De-Vaucouleurs surface brightness radius, $R_{25}$ \protect\citep{RC3}. Open circles are the negative value opacities from NGC 925 and NGC5194-1. }
%\label{f:rsigs}
%\end{figure}

% Figure 4/5 
\begin{figure}
\centering
\includegraphics[angle=0, width=0.5\textwidth]{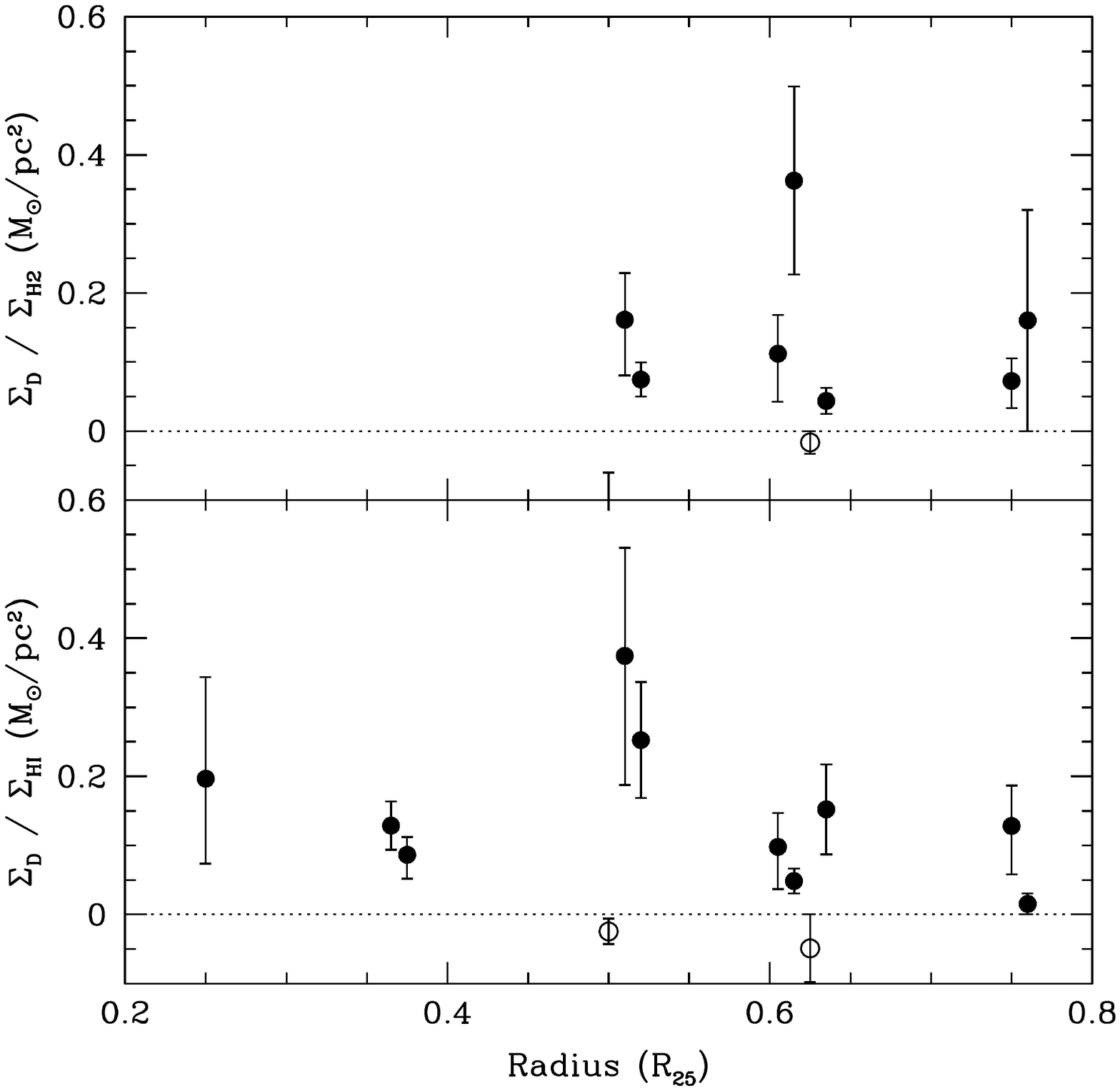}
\caption{The ratio of dust surface density ($\rm \Sigma_{D}$) to either molecular (\h2\ $\rm \Sigma_{H_2}$, top panel) or atomic hydrogen (\hi\ $\rm \Sigma_{HI}$) as a function of galactocentric radius, normalized to the 25 mag/arcsec$^2$ B-band isophotal radius \citep[$R_{25}$ from ][]{RC3}. Open circles are the negative disk opacities from NGC 925 and NGC5194-1. The innermost three points from NGC 3031 and NGC 3621 do not have CO information.}
\label{f:sigsR}
\end{figure}

% holwerda_R_Ratio_Asfm_SigGas.pdf

% Boissier comparison
One explanation for the lack of a relation in Figure \ref{f:sighi-A} is that the measurements were taken at various galactocentric radii (and hence metallicity) in each disk. 
Figure \ref{f:sigsR} plots the ratio between the dust surface density (to facilitate direct comparison) and the two phases of the hydrogen,atomic and molecular in $\rm M_\odot/pc^2$, as a function of radius, scaled to the 25 mag/arcsec$^2$ B-band isophotal radius ($R_{25}$) from \cite{RC3}. 
% HI 
The relation with atomic phase is consistent with a constant fraction of $\rm \Sigma_{D}/\Sigma_{HI} \sim 0.1$ with two exceptions at $R \sim0.5 R_{25}$; NGC 3351 and NGC 3627. Both of these are small \hi\ disks, of which the WFPC2 field covers a large fraction (see Figure \ref{f:himap}), both with prominent spiral arms. NGC 3627 is a member of the Leo triplet and as such may also be a victim of atomic gas stripping or a tidally induced strong spiral pattern. 
% H2
The top panel in \ref{f:sigsR} shows the ratio between dust and molecular surface density, consistent with a constant fraction of $\rm \Sigma_{D}/\Sigma_{H2} \sim 0.75$ with one exception; NGC 3198, which is a very flocculant spiral, which a much lower \hi\ surface density. 
There is little relation between the dust-to-atomic or dust-to-molecular ratio and radius. Exceptions seem to be either strong spiral arm structure, in the case of \hi, or very flocculant, in the case of \h2. 

\begin{figure}
\centering
\includegraphics[angle=0, width=0.5\textwidth]{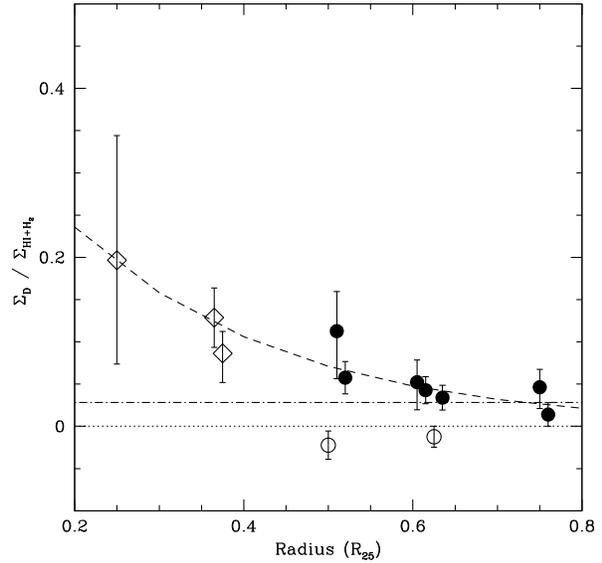}
\caption{ The ratio between implied average dust surface density ($\rm \Sigma_d$), and the total hydrogen surface density ($\rm \Sigma_{HI+H_2}$) as a function of radius. The inner three points (open diamonds) do not have CO data. The dashed-dotted line is the ratio from Bohlin et al. (1978).}
\label{f:ratio}
\end{figure}

%ratio
By combining the surface densities of \hi\ and \h2 into a single hydrogen surface density ($\rm \Sigma_{HI+H2}$), we can now directly compare the total dust-to-gas surface density ratio. In the cases, where no CO observations are available (NGC 3031 and NGC 3621), we use the ratio with \hi\ only.
Figure \ref{f:ratio} shows the dust-to-total-gas ratio as a function of radius. The anomalous ratios of NGC 3351, NGC 3627 and NGC 3198 in Figure \ref{f:sigsR} now fall into line.

If we take the points without CO information (open diamond symbols) at face value (assume no molecular gas), Figure \ref{f:ratio} suggests an exponential decline of dust-to-total gas; 
${\rm \Sigma_d / \Sigma_{HI+H2}} = 0.52 \times e^{- 4.0 {R / R_{25} }}$. A decline of the dust-to-total-gas ratio would be consistent with the relation with metallicity shown in \cite{Leroy11, Sandstrom11}, and with the trends with radius in the recent {\em Spitzer} \citep[e.g.,][]{Munoz-Mateos09b, Bendo10a} and {\em Herschel} results \citep{Pohlen10, Smith10b}. 

However, if we exclude those points without CO information (open diamond symbols) and those with negative SFM measurements (open circles), Figure \ref{f:ratio} is in agreement with a {\em constant} gas-to-dust ratio of $0.043 \pm 0.02$ (weighted mean). One can reasonably expect a much more substantial contribution by the molecular component in the inner disk, which would bring the three points without CO information into line with this constant fraction. This dust-to-total-gas fraction is approximately a factor two above the typical value in the literature \citep[$\sim$0.01-0.03,][]{Smith10b,Leroy11} or the one from \cite{Bohlin78}. The fact that the ratio between dust and total gas surface density is nearly constant points to dust in both the diffuse \hi\ disk as well as in the denser molecular clouds. %  and close to the maximum amount of metals thought to be available for dust grain production.

\section{Discussion}

%% HI alone results
%The two comparisons between disk opacity and \hi\ column density presented above reveal different relations between \hi\ and the dust opacity, depending on the scale over which they are made. When measured on the scale of a WFPC2 field --a sizable part of the disk--, the ratio between disk opacity --the dusty ISM-- and \hi\ shows a gradual decrease in the $A/\Sigma_{\rm HI}$ ratio with radius, as the \hi\ becomes the dominant hydrogen component. 
%% small scale
%At smaller scales, however, opacity and \hi\ column density appear not at all clearly spatially related. This lack of a relation could be the result of our mixing different galaxies and radii, or it may be that the dusty and atomic components are really not in the same locations when viewed at small scales.

% HI OR H2
When compared to either phase of hydrogen in these disks, atomic or molecular, the dust density implied by the disk opacity mostly point to a constant ratio. Exceptions seem to point to a change in gas phase due to the strength of spiral arms in the WFPC2 field-of-view; a strong spiral density wave moves gas into the molecular phase and a flocculant structure into the atomic one. 
A scenario consistent with the density wave origin of spiral structure. In our opinion it illustrates the need for a constraint on both gas phases for a comparison with dust surface density.

% Exponential
% The observed decline with radius of the dust-to-gas ratio agrees with some of the most recent {\em Spitzer} and {\em Herschel} results on dust-to-gas ratio \citep{Munoz-Mateos09b, Bendo10a, Pohlen10, Smith10b}; this decline can be related to gradients in the CO-to-\h2\ conversion factor, or in metallicity \citep[e.g.][]{Leroy11,Sandstrom11}. We note however, that our result is consistent with a {\em constant} dust-to-gas ratio of $\sim$0.05 out to the $R_{25}$, once those data without CO information have been excluded. 

% Dust -- to -- HI+H2 -- to -- Dust surface density comparison.
Our dust-to-total-gas ratio of 0.043 (Figure \ref{f:ratio}) is higher than the values found, for example, in the Local Group spiral galaxies \citep[the Milky Way, M31, and M33 in the case of][]{Leroy11}, or in a single Virgo spiral galaxy with {\em Herschel} \citep[NGC 4501,][]{Smith10b} or the values found by \cite{Magrini11b}. These studies find the values closest to ours in the outskirts of the respective galaxy disks.
% high dust-to-gas ratio
%
% Reasons 
There are several explanations for the high dust-to-gas ratio in our measurements: 
(1) we overestimated the dust surface density, 
(2) a substantial aperture correction of the CO and \hi\ surface densities is needed, 
(3) for large portions of the disk, a different CO-to-\h2\ conversion factor (\xco) is appropriate, and 
(4) a different absorption factor ($\kappa_{abs}$) for a disk average is appropriate.
%
% (1) overestimating the dust from the SFM
First, we are confident that our dust surface densities are unbiased and reasonably accurate because we checked them agains several other observational techniques (Cepheid reddening, occulting galaxy results, {\em Spitzer} FIR SED fits).
Our main assumption is that the dust is in a screen, which is a very rough approximation, especially 
when the probe used is the number of distant objects \citep[i.e, the opacity is also a function of cloud cover][]{Holwerda07b}. 
However, our comparison between dust surface densities from an SED fit and the number count of distant galaxies showed good agreement  \citep{Holwerda07a} (Figure \ref{f:draine}) to within a factor two. 
We note that these SED fits were done without sub-mm information but \cite{Draine07} point out that dust masses can vary with a factor less than 1.5 if the SED of the large grains is done with or without sub-mm information (their Figure 12). The dust surface densities are therefore not likely to be overestimated by more than a factor two. Our comparison with sub-mm fluxes (Figure \ref{f:herschel}, \S \ref{s:herschel}) seems to confirm this.
The SFM estimate of the dust surface density may well be the upper limit of dust in these disks.
\\
%
% (2) underestimating CO and HI fluxes? (aperture issues with weird shaped WFPC2 aperture...)
Secondly, no aperture correction was applied to the CO and \hi\ surface densities. Because the aperture we use to measure the CO and \hi\ fluxes is the odd shape of the WFPC2 camera's field-of-view (Figure \ref{f:himap}), an aperture correction is not straightforward. Yet, we estimate that the aperture correction cannot change the reported average surface brightnesses sufficiently, as the resolution of the observations is substantially smaller than the WFPC2 aperture (Table \ref{t:info}).\\
% 
% (3) averaged over such a big area, the XCO may be underestimating the H2 surface density somewhat as well.
Thirdly, when averaged over a large portion of the disk, which spans a range in density environments, the CO conversion factor (\xco) may well underestimate the total molecular hydrogen surface density, since some molecular clouds the observed CO may be from the  ``skin'' of the GMC and there is not straightforward conversion from CO to \h2\ volume \citep{Wolfire10, Glover11a, Planck-Collaboration11a, Shetty11, Madden11, Feldmann11a,Feldmann12b,  Mac-Low12}. \\
%
% (4)  kappa different? Narayanan11
A fourth option is that the dust absorption factor ($\kappa_{abs}$) is different when averaged over different environments and therefore dust grain properties \citep[e.g.,][]{Narayanan11c,Narayanan12}, although this is likely a secondary effect.
 
If {\em all} our dust surface densities would are all {\em over}-estimated by a factor $\sim$2, or the aperture correction increased the gas surface densities substantially, one may not need to change the \xco\ factor to bring our dust-to-gas ratio in line with recent results from {\em Herschel}. We suspect, however, that the explanation includes a different \xco, when averaged of a large section of the spiral disk and at different galactocentric radii, extending the range in \xco\ values found in Local Group spiral galaxies by \cite{Leroy11}.

% FIGURE 7
\begin{figure}
\centering
\includegraphics[angle=0, width=0.5\textwidth]{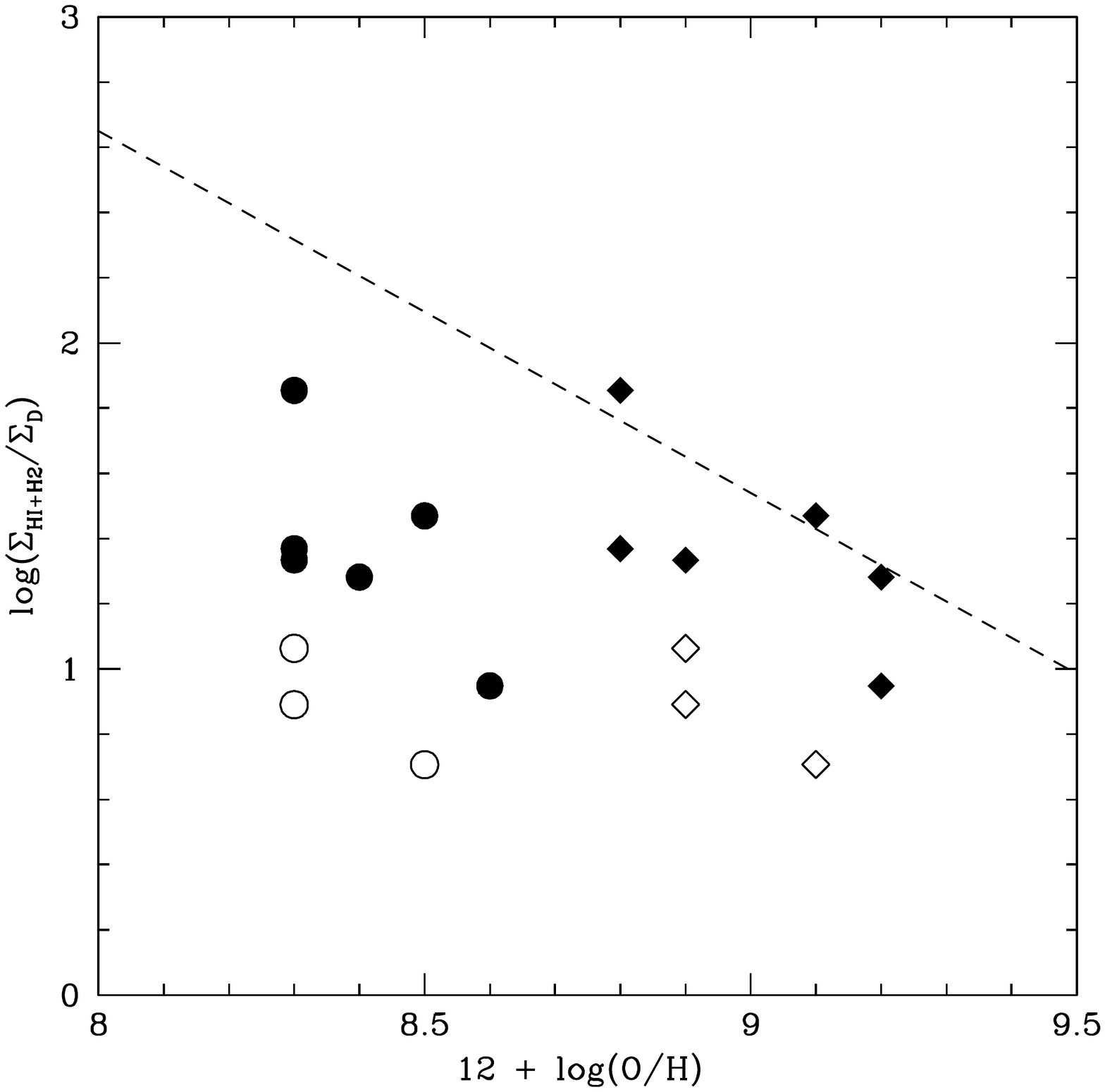}
\caption{The logarithm of the ratio between the total hydrogen surface density ($\rm \Sigma_{HI+H_2}$) and the implied average dust surface density ($\rm \Sigma_d$) as a function of the metallicity ($\rm 12+log(O/H)$), estimated from Figure 7 in \protect\cite{Moustakas10}. 
The circles and diamonds are for the calibration from \protect\cite{KK04} and \protect\cite{PT05} respectively. Open symbols are those points without CO information. There is no metallicity estimate for NGC 3627. We use the gas-to-dust ratio here in order to compare to the relation from \protect\cite{Leroy11}.}
\label{f:metal}
\end{figure}

\subsection{Comparison to Metallicities}
\label{s:metal}

The present consensus is that the dust-to-total-gas ratio depends linearly on the metallicity \citep[see for instance][]{Leroy11}. 
Fortunately, uniformly determined metallicity gradients for the SINGS\footnote{Spitzer Infrared Nearby Galaxy Survey \protect\citep{SINGS}.}, and hence 
THINGS and HERACLES, galaxies are presented in \cite{Moustakas10}. Only NGC 3627 does not have metallicity information.
Starting from their linear relation for the radial dependence of metallicity in each galaxy (their Figure 7), 
we can obtain an estimate of the metallicity for each of our WFPC2 fields. They present two different estimates 
of metallicity ($\rm log(O/H)$), with either the theoretical calibration from \cite{KK04} or the empirical one from 
\cite{PT05} (see Table \ref{t:info}). \cite{Moustakas10} note that, until the calibration issues are resolved, one 
should either average the metallicity estimates based on either calibration or use both separately.  
We will use both calibrations separately for comparison and the total-gas-to-dust ratio to facilitate a direct comparison to Figure 6 in \cite{Leroy11}. 
We note that since our WFPC2 fields were placed with crowding issues in mind, our coverage of galactocentric 
radii (and hence metallicities) is not very large. 

Figure \ref{f:metal} shows the logarithm of the total-gas-to-dust ratio as a function of metallicity, using either of the two calibrations. 
Our points lie lower than the linear relation from  \cite{Leroy11} for the gas-to-dust ratio with metallicity, 
not unexpectedly as we already established that our dust-to-gas values are higher than those previously reported.

However, using the calibration from \cite{PT05}, and discarding those points that lack CO information, 
there is a reasonable agreement with the relation from \cite{Leroy11}.

\section{Conclusions}
\label{s:concl}

To conclude our ``Opacity of spiral disks" project, we have compared the opacity of spiral galaxies, and the hence dust surface density to the surface densities of hydrogen, both atomic and molecular, the original goal of our project. We conclude from this comparison:

\begin{enumerate}
\item The disk opacity scales with the {\em Herschel-SPIRE} 250 $\mu$m surface brightness (Figure \ref{f:herschel}), confirming our assertion that opacity scales with dust surface density to first order.
\item There is little relation between the \hi\ column density and where a distant galaxy was identified in these fields (Figure \ref{f:na}).
\item Averaged over a WFPC2 field, there is only a weak link between disk opacity (or dust surface density) and gas surface density, either atomic, molecular or total (Figure \ref{f:sighi-A}), pointing to third factor; radius or metallicity.
\item The dust-to-\hi\ or dust-to-\h2\ relations with galactocentric radius are both relatively constant (Figure \ref{f:sigsR}), but the exceptions point to the role of spiral structure in the dominant gas phase of the ISM.
\item The dust-to-{\em total}-gas ratio is close to constant for all our fields $\rm \Sigma_{HI+H_2} = 0.043 \pm 0.024$ (Figure \ref{f:ratio}). This higher value can, in our opinion, be attributed to a different conversion to dust surface density or the CO-to-\h2 conversion factor (\xco) for such large sections of disks. 
\item Compared to the relation between total-gas-to-dust and metallicity from \cite{Leroy11}, our results are reasonably consistent, provided one uses the \cite{PT05} calibration of the metallicities of \cite{Moustakas10} (Figure \ref{f:metal}). 
\end{enumerate}

Future use of the number of distant galaxies identified through a foreground spiral disk as a probe of dust is critically limited by cosmic variance \citep{Gonzalez03, Holwerda05e} but 
its optimal application will be on a {\em single} large {\em HST} mosaic of a nearby face-on spiral (e.g., M81 or M101), which will most likely the last contribution of this unique approach to the issue of the dust content of spiral disks.

\section*{Acknowledgements}

We acknowledge the THINGS collaboration for the publication of their \hi\  surface density maps, based on their Very Large Array radio observations and would like to thank the HERACLES collaboration for making their surface density maps available early. The authors would like to thank F. Walther and S-L. Blyth for useful discussions and feedback. We thank the anonymous referee for his or her excellent report and extraordinary effort.

We acknowledge support from HST Archive grants AR-10662 and AR-10663 and from the National Research Foundation of South Africa. The work of W.J.G. de Blok is based upon research supported by the South African Research Chairs Initiative of the Department of Science and Technology and the National Research Foundation. Antoine Bouchard acknowledges the financial support from the South African Square Kilometre Array Project. R. A. Gonz\'alez-L\'opezlira acknowledges support from  DGAPA (UNAM) grant IN118110.

The National Radio Astronomy Observatory is a facility of the National Science Foundation operated under cooperative agreement by Associated Universities, Inc. Based on observations made with the NASA/ESA Hubble Space Telescope, which is a collaboration between the Space Telescope Science Institute 
(STScI/ NASA), the Space Telescope European Coordinating Facility (ST-ECF/ ESA), and the Canadian Astronomy Data Centre (CADC/NRC/CSA). The Hubble data presented in this paper were obtained from the Multimission Archive at the Space Telescope Science Institute (MAST). STScI is operated by the Association of Universities for Research in Astronomy, Inc., under NASA contract NAS5-26555. Support for MAST for non-HST data is provided by the NASA Office of Space Science via grant NAG5-7584, and by other grants and contracts.

This research has made use of the NASA/IPAC Extragalactic Database (NED) which is operated by the Jet Propulsion Laboratory, California Institute of Technology, under contract with the National Aeronautics and Space Administration.
This research has made use of NASA's Astrophysics Data System.

%\bibliographystyle{aa} 
%\bibliography{/Users/bholwerd/Desktop/Science/Bib/Bibliography}

\begin{thebibliography}{123}
\expandafter\ifx\csname natexlab\endcsname\relax\def\natexlab#1{#1}\fi

\bibitem[{{Allen} {et~al.}(1986){Allen}, {Atherton}, \& {Tilanus}}]{Allen86}
{Allen}, R.~J., {Atherton}, P.~D., \& {Tilanus}, R.~P.~J. 1986, \nat, 319, 296

\bibitem[{{Allen} {et~al.}(2004){Allen}, {Heaton}, \& {Kaufman}}]{Allen04}
{Allen}, R.~J., {Heaton}, H.~I., \& {Kaufman}, M.~J. 2004, \apj, 608, 314

\bibitem[{{Allen} {et~al.}(1997){Allen}, {Knapen}, {Bohlin}, \&
  {Stecher}}]{Allen97}
{Allen}, R.~J., {Knapen}, J.~H., {Bohlin}, R., \& {Stecher}, T.~P. 1997, \apj,
  487, 171

\bibitem[{{Andredakis} \& {van der Kruit}(1992)}]{Andredakis92}
{Andredakis}, Y.~C. \& {van der Kruit}, P.~C. 1992, \aap, 265, 396

\bibitem[{{Bell} {et~al.}(2003){Bell}, {Baugh}, {Cole}, {Frenk}, \&
  {Lacey}}]{Bell03c}
{Bell}, E.~F., {Baugh}, C.~M., {Cole}, S., {Frenk}, C.~S., \& {Lacey}, C.~G.
  2003, \mnras, 343, 367

\bibitem[{{Bendo} {et~al.}(2008){Bendo}, {Draine}, {Engelbracht}, {Helou},
  {Thornley}, {Bot}, {Buckalew}, {Calzetti}, {Dale}, {Hollenbach}, {Li}, \&
  {Moustakas}}]{Bendo08}
{Bendo}, G.~J., {Draine}, B.~T., {Engelbracht}, C.~W., {et~al.} 2008, \mnras,
  389, 629

\bibitem[{{Bendo} {et~al.}(2010{\natexlab{a}}){Bendo}, {Wilson}, {Pohlen},
  {Sauvage}, {Auld}, {Baes}, {Barlow}, {Bock}, {Boselli}, {Bradford}, {Buat},
  {Castro-Rodriguez}, {Chanial}, {Charlot}, {Ciesla}, {Clements}, {Cooray},
  {Cormier}, {Cortese}, {Davies}, {Dwek}, {Eales}, {Elbaz}, {Galametz},
  {Galliano}, {Gear}, {Glenn}, {Gomez}, {Griffin}, {Hony}, {Isaak}, {Levenson},
  {Lu}, {Madden}, {O'Halloran}, {Okumura}, {Oliver}, {Page}, {Panuzzo},
  {Papageorgiou}, {Parkin}, {Perez-Fournon}, {Rangwala}, {Rigby}, {Roussel},
  {Rykala}, {Sacchi}, {Schulz}, {Schirm}, {Smith}, {Spinoglio}, {Stevens},
  {Sundar}, {Symeonidis}, {Trichas}, {Vaccari}, {Vigroux}, {Wozniak}, {Wright},
  \& {Zeilinger}}]{Bendo10a}
{Bendo}, G.~J., {Wilson}, C.~D., {Pohlen}, M., {et~al.} 2010{\natexlab{a}},
  \aap, 518, L65+

\bibitem[{{Bendo} {et~al.}(2010{\natexlab{b}}){Bendo}, {Wilson}, {Warren},
  {Brinks}, {Butner}, {Chanial}, {Clements}, {Courteau}, {Irwin}, {Israel},
  {Knapen}, {Leech}, {Matthews}, {M{\"u}hle}, {Petitpas}, {Serjeant}, {Tan},
  {Tilanus}, {Usero}, {Vaccari}, {van der Werf}, {Vlahakis}, {Wiegert}, \&
  {Zhu}}]{Bendo10b}
{Bendo}, G.~J., {Wilson}, C.~D., {Warren}, B.~E., {et~al.} 2010{\natexlab{b}},
  \mnras, 402, 1409

\bibitem[{{Berlind} {et~al.}(1997){Berlind}, {Quillen}, {Pogge}, \&
  {Sellgren}}]{Berlind97}
{Berlind}, A.~A., {Quillen}, A.~C., {Pogge}, R.~W., \& {Sellgren}, K. 1997,
  \aj, 114, 107

\bibitem[{{Bigiel} {et~al.}(2008){Bigiel}, {Leroy}, {Walter}, {Brinks}, {de
  Blok}, {Madore}, \& {Thornley}}]{Bigiel08}
{Bigiel}, F., {Leroy}, A., {Walter}, F., {et~al.} 2008, \aj, 136, 2846

\bibitem[{{Bohlin} {et~al.}(1978){Bohlin}, {Savage}, \& {Drake}}]{Bohlin78}
{Bohlin}, R.~C., {Savage}, B.~D., \& {Drake}, J.~F. 1978, \apj, 224, 132

\bibitem[{{Boissier} {et~al.}(2004){Boissier}, {Boselli}, {Buat}, {Donas}, \&
  {Milliard}}]{Boissier04}
{Boissier}, S., {Boselli}, A., {Buat}, V., {Donas}, J., \& {Milliard}, B. 2004,
  \aap, 424, 465

\bibitem[{{Boissier} {et~al.}(2007){Boissier}, {Gil de Paz}, {Boselli},
  {Madore}, {Buat}, {Cortese}, {Burgarella}, {Mu{\~n}oz-Mateos}, {Barlow},
  {Forster}, {Friedman}, {Martin}, {Morrissey}, {Neff}, {Schiminovich},
  {Seibert}, {Small}, {Wyder}, {Bianchi}, {Donas}, {Heckman}, {Lee},
  {Milliard}, {Rich}, {Szalay}, {Welsh}, \& {Yi}}]{Boissier07}
{Boissier}, S., {Gil de Paz}, A., {Boselli}, A., {et~al.} 2007, \apjs, 173, 524

\bibitem[{{Boissier} {et~al.}(2005){Boissier}, {Gil de Paz}, {Madore},
  {Boselli}, {Buat}, {Burgarella}, {Friedman}, {Barlow}, {Bianchi}, {Byun},
  {Donas}, {Forster}, {Heckman}, {Jelinsky}, {Lee}, {Malina}, {Martin},
  {Milliard}, {Morrissey}, {Neff}, {Rich}, {Schiminovich}, {Siegmund}, {Small},
  {Szalay}, {Welsh}, \& {Wyder}}]{Boissier05}
{Boissier}, S., {Gil de Paz}, A., {Madore}, B.~F., {et~al.} 2005, \apjl, 619,
  L83

\bibitem[{{Boselli} {et~al.}(2010){Boselli}, {Eales}, {Cortese}, {Bendo},
  {Chanial}, {Buat}, {Davies}, {Auld}, {Rigby}, {Baes}, {Barlow}, {Bock},
  {Bradford}, {Castro-Rodriguez}, {Charlot}, {Clements}, {Cormier}, {Dwek},
  {Elbaz}, {Galametz}, {Galliano}, {Gear}, {Glenn}, {Gomez}, {Griffin}, {Hony},
  {Isaak}, {Levenson}, {Lu}, {Madden}, {O'Halloran}, {Okamura}, {Oliver},
  {Page}, {Panuzzo}, {Papageorgiou}, {Parkin}, {Perez-Fournon}, {Pohlen},
  {Rangwala}, {Roussel}, {Rykala}, {Sacchi}, {Sauvage}, {Schulz}, {Schirm},
  {Smith}, {Spinoglio}, {Stevens}, {Symeonidis}, {Vaccari}, {Vigroux},
  {Wilson}, {Wozniak}, {Wright}, \& {Zeilinger}}]{Boselli10}
{Boselli}, A., {Eales}, S., {Cortese}, L., {et~al.} 2010, \pasp, 122, 261

\bibitem[{{Braine} {et~al.}(2010){Braine}, {Gratier}, {Kramer}, {Xilouris},
  {Rosolowsky}, {Buchbender}, {Boquien}, {Calzetti}, {Quintana-Lacaci},
  {Tabatabaei}, {Verley}, {Israel}, {van der Tak}, {Aalto}, {Combes},
  {Garcia-Burillo}, {Gonzalez}, {Henkel}, {Koribalski}, {Mookerjea}, {Roellig},
  {Schuster}, {Rela{\~n}o}, {Bertoldi}, {van der Werf}, \&
  {Wiedner}}]{Braine10}
{Braine}, J., {Gratier}, P., {Kramer}, C., {et~al.} 2010, \aap, 518, L69+

\bibitem[{{Buat} {et~al.}(2002){Buat}, {Boselli}, {Gavazzi}, \&
  {Bonfanti}}]{Buat02}
{Buat}, V., {Boselli}, A., {Gavazzi}, G., \& {Bonfanti}, C. 2002, \aap, 383,
  801

\bibitem[{{Calzetti}(2001)}]{Calzetti01}
{Calzetti}, D. 2001, \pasp, 113, 1449

\bibitem[{{Calzetti} {et~al.}(2005){Calzetti}, {Kennicutt}, {Bianchi},
  {Thilker}, {Dale}, {Engelbracht}, {Leitherer}, {Meyer}, {Sosey}, {Mutchler},
  {Regan}, {Thornley}, {Armus}, {Bendo}, {Boissier}, {Boselli}, {Draine},
  {Gordon}, {Helou}, {Hollenbach}, {Kewley}, {Madore}, {Martin}, {Murphy},
  {Rieke}, {Rieke}, {Roussel}, {Sheth}, {Smith}, {Walter}, {White}, {Yi},
  {Scoville}, {Polletta}, \& {Lindler}}]{Calzetti05}
{Calzetti}, D., {Kennicutt}, R.~C., {Bianchi}, L., {et~al.} 2005, \apj, 633,
  871

\bibitem[{{Catinella} {et~al.}(2010){Catinella}, {Schiminovich}, {Kauffmann},
  {Fabello}, {Wang}, {Hummels}, {Lemonias}, {Moran}, {Wu}, {Giovanelli},
  {Haynes}, {Heckman}, {Basu-Zych}, {Blanton}, {Brinchmann}, {Budav{\'a}ri},
  {Gon{\c c}alves}, {Johnson}, {Kennicutt}, {Madore}, {Martin}, {Rich},
  {Tacconi}, {Thilker}, {Wild}, \& {Wyder}}]{Catinella10}
{Catinella}, B., {Schiminovich}, D., {Kauffmann}, G., {et~al.} 2010, \mnras,
  403, 683

\bibitem[{{Cazaux} \& {Tielens}(2004)}]{Cazaux04b}
{Cazaux}, S. \& {Tielens}, A.~G.~G.~M. 2004, \apj, 604, 222

\bibitem[{{Combes} {et~al.}(2012){Combes}, {Boquien}, {Kramer}, {Xilouris},
  {Bertoldi}, {Braine}, {Buchbender}, {Calzetti}, {Gratier}, {Israel},
  {Koribalski}, {Lord}, {Quintana-Lacaci}, {Rela{\~n}o}, {R{\"o}llig},
  {Stacey}, {Tabatabaei}, {Tilanus}, {van der Tak}, {van der Werf}, \&
  {Verley}}]{Combes12}
{Combes}, F., {Boquien}, M., {Kramer}, C., {et~al.} 2012, \aap, 539, A67

\bibitem[{{Cortese} {et~al.}(2006){Cortese}, {Boselli}, {Buat}, {Gavazzi},
  {Boissier}, {Gil de Paz}, {Seibert}, {Madore}, \& {Martin}}]{Cortese06a}
{Cortese}, L., {Boselli}, A., {Buat}, V., {et~al.} 2006, \apj, 637, 242

\bibitem[{{Cortese} {et~al.}(2010){Cortese}, {Davies}, {Pohlen}, {Baes},
  {Bendo}, {Bianchi}, {Boselli}, {de Looze}, {Fritz}, {Verstappen}, {Bomans},
  {Clemens}, {Corbelli}, {Dariush}, {di Serego Alighieri}, {Fadda},
  {Garcia-Appadoo}, {Gavazzi}, {Giovanardi}, {Grossi}, {Hughes}, {Hunt},
  {Jones}, {Madden}, {Pierini}, {Sabatini}, {Smith}, {Vlahakis}, {Xilouris}, \&
  {Zibetti}}]{Cortese10}
{Cortese}, L., {Davies}, J.~I., {Pohlen}, M., {et~al.} 2010, \aap, 518, L49+

\bibitem[{{Cuillandre} {et~al.}(2001){Cuillandre}, {Lequeux}, {Allen},
  {Mellier}, \& {Bertin}}]{Cuillandre01}
{Cuillandre}, J., {Lequeux}, J., {Allen}, R.~J., {Mellier}, Y., \& {Bertin}, E.
  2001, \apj, 554, 190

\bibitem[{{da Cunha} {et~al.}(2008){da Cunha}, {Charlot}, \&
  {Elbaz}}]{da-Cunha08}
{da Cunha}, E., {Charlot}, S., \& {Elbaz}, D. 2008, \mnras, 388, 1595

\bibitem[{{Dale} {et~al.}(2012){Dale}, {Aniano}, {Engelbracht}, {Hinz},
  {Krause}, {Montiel}, {Roussel}, {Appleton}, {Armus}, {Beir{\~a}o}, {Bolatto},
  {Brandl}, {Calzetti}, {Crocker}, {Croxall}, {Draine}, {Galametz}, {Gordon},
  {Groves}, {Hao}, {Helou}, {Hunt}, {Johnson}, {Kennicutt}, {Koda}, {Leroy},
  {Li}, {Meidt}, {Miller}, {Murphy}, {Rahman}, {Rix}, {Sandstrom}, {Sauvage},
  {Schinnerer}, {Skibba}, {Smith}, {Tabatabaei}, {Walter}, {Wilson}, {Wolfire},
  \& {Zibetti}}]{Dale12}
{Dale}, D.~A., {Aniano}, G., {Engelbracht}, C.~W., {et~al.} 2012, \apj, 745, 95

\bibitem[{{de Vaucouleurs} {et~al.}(1991){de Vaucouleurs}, {de Vaucouleurs},
  {Corwin}, {Buta}, {Paturel}, \& {Fouque}}]{RC3}
{de Vaucouleurs}, G., {de Vaucouleurs}, A., {Corwin}, H.~G., {et~al.} 1991,
  {Third Reference Catalogue of Bright Galaxies} (Volume 1-3, XII, 2069 pp.~7
  figs..~ Springer-Verlag Berlin Heidelberg New York)

\bibitem[{{Domingue} {et~al.}(1999){Domingue}, {Keel}, {Ryder}, \&
  {White}}]{kw99a}
{Domingue}, D.~L., {Keel}, W.~C., {Ryder}, S.~D., \& {White}, III, R.~E. 1999,
  \aj, 118, 1542

\bibitem[{{Domingue} {et~al.}(2000){Domingue}, {Keel}, \& {White}}]{kw00b}
{Domingue}, D.~L., {Keel}, W.~C., \& {White}, III, R.~E. 2000, \apj, 545, 171

\bibitem[{{Draine}(2003)}]{Draine03}
{Draine}, B.~T. 2003, \araa, 41, 241

\bibitem[{{Draine} {et~al.}(2007){Draine}, {Dale}, {Bendo}, {Gordon}, {Smith},
  {Armus}, {Engelbracht}, {Helou}, {Kennicutt}, {Li}, {Roussel}, {Walter},
  {Calzetti}, {Moustakas}, {Murphy}, {Rieke}, {Bot}, {Hollenbach}, {Sheth}, \&
  {Teplitz}}]{Draine07}
{Draine}, B.~T., {Dale}, D.~A., {Bendo}, G., {et~al.} 2007, \apj, 663, 866

\bibitem[{{Dutra} {et~al.}(2001){Dutra}, {Bica}, {Clari{\'a}}, {Piatti}, \&
  {Ahumada}}]{Dutra01}
{Dutra}, C.~M., {Bica}, E., {Clari{\'a}}, J.~J., {Piatti}, A.~E., \& {Ahumada},
  A.~V. 2001, \aap, 371, 895

\bibitem[{{Eales} {et~al.}(2010){Eales}, {Smith}, {Wilson}, {Bendo}, {Cortese},
  {Pohlen}, {Boselli}, {Gomez}, {Auld}, {Baes}, {Barlow}, {Bock}, {Bradford},
  {Buat}, {Castro-Rodr{\'{\i}}guez}, {Chanial}, {Charlot}, {Ciesla},
  {Clements}, {Cooray}, {Cormier}, {Davies}, {Dwek}, {Elbaz}, {Galametz},
  {Galliano}, {Gear}, {Glenn}, {Griffin}, {Hony}, {Isaak}, {Levenson}, {Lu},
  {Madden}, {O'Halloran}, {Okumura}, {Oliver}, {Page}, {Panuzzo},
  {Papageorgiou}, {Parkin}, {P{\'e}rez-Fournon}, {Rangwala}, {Rigby},
  {Roussel}, {Rykala}, {Sacchi}, {Sauvage}, {Schulz}, {Schirm}, {Spinoglio},
  {Srinivasan}, {Stevens}, {Symeonidis}, {Trichas}, {Vaccari}, {Vigroux},
  {Wozniak}, {Wright}, \& {Zeilinger}}]{Eales10}
{Eales}, S.~A., {Smith}, M.~W.~L., {Wilson}, C.~D., {et~al.} 2010, \aap, 518,
  L62+

\bibitem[{{Elmegreen} {et~al.}(2001){Elmegreen}, {Kaufman}, {Elmegreen},
  {Brinks}, {Struck}, {Klari{\'c}}, \& {Thomasson}}]{Elmegreen01}
{Elmegreen}, D.~M., {Kaufman}, M., {Elmegreen}, B.~G., {et~al.} 2001, \aj, 121,
  182

\bibitem[{{Engelbracht} {et~al.}(2010){Engelbracht}, {Hunt}, {Skibba}, {Hinz},
  {Calzetti}, {Gordon}, {Roussel}, {Crocker}, {Misselt}, {Bolatto},
  {Kennicutt}, {Appleton}, {Armus}, {Beir{\~a}o}, {Brandl}, {Croxall}, {Dale},
  {Draine}, {Dumas}, {Gil de Paz}, {Groves}, {Hao}, {Johnson}, {Koda},
  {Krause}, {Leroy}, {Meidt}, {Murphy}, {Rahman}, {Rix}, {Sandstrom},
  {Sauvage}, {Schinnerer}, {Smith}, {Srinivasan}, {Vigroux}, {Walter},
  {Warren}, {Wilson}, {Wolfire}, \& {Zibetti}}]{Engelbracht10}
{Engelbracht}, C.~W., {Hunt}, L.~K., {Skibba}, R.~A., {et~al.} 2010, \aap, 518,
  L56+

\bibitem[{{Fabello} {et~al.}(2011){Fabello}, {Catinella}, {Giovanelli},
  {Kauffmann}, {Haynes}, {Heckman}, \& {Schiminovich}}]{Fabello11a}
{Fabello}, S., {Catinella}, B., {Giovanelli}, R., {et~al.} 2011, \mnras, 411,
  993

\bibitem[{{Fall} \& {Efstathiou}(1980)}]{Fall80}
{Fall}, S.~M. \& {Efstathiou}, G. 1980, \mnras, 193, 189

\bibitem[{{Feldmann} {et~al.}(2011{\natexlab{a}}){Feldmann}, {Gnedin}, \&
  {Kravtsov}}]{Feldmann11a}
{Feldmann}, R., {Gnedin}, N.~Y., \& {Kravtsov}, A.~V. 2011{\natexlab{a}}, \apj,
  732, 115

\bibitem[{{Feldmann} {et~al.}(2011{\natexlab{b}}){Feldmann}, {Gnedin}, \&
  {Kravtsov}}]{Feldmann11b}
{Feldmann}, R., {Gnedin}, N.~Y., \& {Kravtsov}, A.~V. 2011{\natexlab{b}}, ArXiv
  e-prints

\bibitem[{{Foyle} {et~al.}(2012){Foyle}, {Wilson}, {Mentuch}, {Bendo},
  {Dariush}, {Parkin}, {Pohlen}, {Sauvage}, {Smith}, {Roussel}, {Baes},
  {Boquien}, {Boselli}, {Clements}, {Cooray}, {Davies}, {Eales}, {Madden},
  {Page}, \& {Spinoglio}}]{Foyle12}
{Foyle}, K., {Wilson}, C.~D., {Mentuch}, E., {et~al.} 2012, ArXiv e-prints

\bibitem[{{Freedman} {et~al.}(2001){Freedman}, {Madore}, {Gibson}, {Ferrarese},
  {Kelson}, {Sakai}, {Mould}, {Kennicutt}, {Ford}, {Graham}, {Huchra},
  {Hughes}, {Illingworth}, {Macri}, \& {Stetson}}]{KeyProject}
{Freedman}, W.~L., {Madore}, B.~F., {Gibson}, B.~K., {et~al.} 2001, \apj, 553,
  47

\bibitem[{{Galametz} {et~al.}(2012){Galametz}, {Kennicutt}, {Albrecht},
  {Aniano}, {Armus}, {Bertoldi}, {Calzetti}, {Crocker}, {Croxall}, {Dale},
  {Donovan Meyer}, {Draine}, {Engelbracht}, {Hinz}, {Roussel}, {Skibba},
  {Tabatabaei}, {Walter}, {Weiss}, {Wilson}, \& {Wolfire}}]{Galametz12}
{Galametz}, M., {Kennicutt}, R.~C., {Albrecht}, M., {et~al.} 2012, ArXiv
  e-prints

\bibitem[{{Galliano} {et~al.}(2011){Galliano}, {Hony}, {Bernard}, {Bot},
  {Madden}, {Roman-Duval}, {Galametz}, {Li}, {Meixner}, {Engelbracht},
  {Lebouteiller}, {Misselt}, {Montiel}, {Panuzzo}, {Reach}, \&
  {Skibba}}]{Galliano11}
{Galliano}, F., {Hony}, S., {Bernard}, J.~., {et~al.} 2011, ArXiv e-prints

\bibitem[{{Glover} \& {Mac Low}(2011)}]{Glover11a}
{Glover}, S.~C.~O. \& {Mac Low}, M. 2011, \mnras, 412, 337

\bibitem[{{Gonz{\'a}lez} {et~al.}(1998){Gonz{\'a}lez}, {Allen}, {Dirsch},
  {Ferguson}, {Calzetti}, \& {Panagia}}]{Gonzalez98}
{Gonz{\'a}lez}, R.~A., {Allen}, R.~J., {Dirsch}, B., {et~al.} 1998, \apj, 506,
  152

\bibitem[{{Gonz{\'a}lez} {et~al.}(2003){Gonz{\'a}lez}, {Loinard}, {Allen}, \&
  {Muller}}]{Gonzalez03}
{Gonz{\'a}lez}, R.~A., {Loinard}, L., {Allen}, R.~J., \& {Muller}, S. 2003,
  \aj, 125, 1182

\bibitem[{{Gordon} {et~al.}(2008){Gordon}, {Engelbracht}, {Rieke}, {Misselt},
  {Smith}, \& {Kennicutt}}]{Gordon08}
{Gordon}, K.~D., {Engelbracht}, C.~W., {Rieke}, G.~H., {et~al.} 2008, \apj,
  682, 336

\bibitem[{{Gordon} {et~al.}(2010){Gordon}, {Galliano}, {Hony}, {Bernard},
  {Bolatto}, {Bot}, {Engelbracht}, {Hughes}, {Israel}, {Kemper}, {Kim}, {Li},
  {Madden}, {Matsuura}, {Meixner}, {Misselt}, {Okumura}, {Panuzzo}, {Rubio},
  {Reach}, {Roman-Duval}, {Sauvage}, {Skibba}, \& {Tielens}}]{Gordon10}
{Gordon}, K.~D., {Galliano}, F., {Hony}, S., {et~al.} 2010, \aap, 518, L89+

\bibitem[{{Griffin} {et~al.}(2010){Griffin}, {Abergel}, {Abreu}, {Ade},
  {Andr{\'e}}, {Augueres}, {Babbedge}, {Bae}, {Baillie}, {Baluteau}, {Barlow},
  {Bendo}, {Benielli}, {Bock}, {Bonhomme}, {Brisbin}, {Brockley-Blatt},
  {Caldwell}, {Cara}, {Castro-Rodriguez}, {Cerulli}, {Chanial}, {Chen},
  {Clark}, {Clements}, {Clerc}, {Coker}, {Communal}, {Conversi}, {Cox},
  {Crumb}, {Cunningham}, {Daly}, {Davis}, {de Antoni}, {Delderfield}, {Devin},
  {di Giorgio}, {Didschuns}, {Dohlen}, {Donati}, {Dowell}, {Dowell}, {Duband},
  {Dumaye}, {Emery}, {Ferlet}, {Ferrand}, {Fontignie}, {Fox}, {Franceschini},
  {Frerking}, {Fulton}, {Garcia}, {Gastaud}, {Gear}, {Glenn}, {Goizel},
  {Griffin}, {Grundy}, {Guest}, {Guillemet}, {Hargrave}, {Harwit}, {Hastings},
  {Hatziminaoglou}, {Herman}, {Hinde}, {Hristov}, {Huang}, {Imhof}, {Isaak},
  {Israelsson}, {Ivison}, {Jennings}, {Kiernan}, {King}, {Lange}, {Latter},
  {Laurent}, {Laurent}, {Leeks}, {Lellouch}, {Levenson}, {Li}, {Li},
  {Lilienthal}, {Lim}, {Liu}, {Lu}, {Madden}, {Mainetti}, {Marliani}, {McKay},
  {Mercier}, {Molinari}, {Morris}, {Moseley}, {Mulder}, {Mur}, {Naylor},
  {Nguyen}, {O'Halloran}, {Oliver}, {Olofsson}, {Olofsson}, {Orfei}, {Page},
  {Pain}, {Panuzzo}, {Papageorgiou}, {Parks}, {Parr-Burman}, {Pearce},
  {Pearson}, {P{\'e}rez-Fournon}, {Pinsard}, {Pisano}, {Podosek}, {Pohlen},
  {Polehampton}, {Pouliquen}, {Rigopoulou}, {Rizzo}, {Roseboom}, {Roussel},
  {Rowan-Robinson}, {Rownd}, {Saraceno}, {Sauvage}, {Savage}, {Savini},
  {Sawyer}, {Scharmberg}, {Schmitt}, {Schneider}, {Schulz}, {Schwartz},
  {Shafer}, {Shupe}, {Sibthorpe}, {Sidher}, {Smith}, {Smith}, {Smith},
  {Spencer}, {Stobie}, {Sudiwala}, {Sukhatme}, {Surace}, {Stevens}, {Swinyard},
  {Trichas}, {Tourette}, {Triou}, {Tseng}, {Tucker}, {Turner}, {Vaccari},
  {Valtchanov}, {Vigroux}, {Virique}, {Voellmer}, {Walker}, {Ward}, {Waskett},
  {Weilert}, {Wesson}, {White}, {Whitehouse}, {Wilson}, {Winter}, {Woodcraft},
  {Wright}, {Xu}, {Zavagno}, {Zemcov}, {Zhang}, \& {Zonca}}]{Griffin10}
{Griffin}, M.~J., {Abergel}, A., {Abreu}, A., {et~al.} 2010, \aap, 518, L3+

\bibitem[{{Gurwell} \& {Hodge}(1990)}]{Gurwell90}
{Gurwell}, M. \& {Hodge}, P. 1990, \pasp, 102, 849

\bibitem[{{Heiner} {et~al.}(2008{\natexlab{a}}){Heiner}, {Allen}, {Emonts}, \&
  {van der Kruit}}]{Heiner08a}
{Heiner}, J.~S., {Allen}, R.~J., {Emonts}, B.~H.~C., \& {van der Kruit}, P.~C.
  2008{\natexlab{a}}, \apj, 673, 798

\bibitem[{{Heiner} {et~al.}(2009){Heiner}, {Allen}, \& {van der
  Kruit}}]{Heiner09}
{Heiner}, J.~S., {Allen}, R.~J., \& {van der Kruit}, P.~C. 2009, in The
  Evolving ISM in the Milky Way and Nearby Galaxies

\bibitem[{{Heiner} {et~al.}(2010){Heiner}, {Allen}, \& {van der
  Kruit}}]{Heiner10}
{Heiner}, J.~S., {Allen}, R.~J., \& {van der Kruit}, P.~C. 2010, \apj, 719,
  1244

\bibitem[{{Heiner} {et~al.}(2008{\natexlab{b}}){Heiner}, {Allen}, {Wong}, \&
  {van der Kruit}}]{Heiner08b}
{Heiner}, J.~S., {Allen}, R.~J., {Wong}, O.~I., \& {van der Kruit}, P.~C.
  2008{\natexlab{b}}, \aap, 489, 533

\bibitem[{{Hodge}(1974)}]{Hodge74}
{Hodge}, P.~W. 1974, \apj, 192, 21

\bibitem[{{Hodge} \& {Snow}(1975)}]{Hodge75}
{Hodge}, P.~W. \& {Snow}, T.~P. 1975, \aj, 80, 9

\bibitem[{{Holwerda}(2005{\natexlab{a}})}]{Holwerda05}
{Holwerda}, B.~W. 2005{\natexlab{a}}, PhD thesis, Proefschrift,
  Rijksuniversiteit Groningen, 2005

\bibitem[{{Holwerda}(2005{\natexlab{b}})}]{mythesis}
{Holwerda}, B.~W. 2005{\natexlab{b}}, PhD thesis, Proefschrift,
  Rijksuniversiteit Groningen, 2005

\bibitem[{{Holwerda} {et~al.}(2007{\natexlab{a}}){Holwerda}, {Draine},
  {Gordon}, {Gonz\'alez}, {Calzetti}, {Thornley}, {Buckalew}, {Allen}, \& {van
  der Kruit}}]{Holwerda07a}
{Holwerda}, B.~W., {Draine}, B., {Gordon}, K.~D., {et~al.} 2007{\natexlab{a}},
  \aj, 134, 2226

\bibitem[{{Holwerda} {et~al.}(2005{\natexlab{a}}){Holwerda}, {Gonz\'alez},
  {Allen}, \& {van der Kruit}}]{Holwerda05a}
{Holwerda}, B.~W., {Gonz\'alez}, R.~A., {Allen}, R.~J., \& {van der Kruit},
  P.~C. 2005{\natexlab{a}}, \aj, 129, 1381

\bibitem[{{Holwerda} {et~al.}(2005{\natexlab{b}}){Holwerda}, {Gonz\'alez},
  {Allen}, \& {van der Kruit}}]{Holwerda05b}
{Holwerda}, B.~W., {Gonz\'alez}, R.~A., {Allen}, R.~J., \& {van der Kruit},
  P.~C. 2005{\natexlab{b}}, \aj, 129, 1396

\bibitem[{{Holwerda} {et~al.}(2005{\natexlab{c}}){Holwerda}, {Gonz\'alez},
  {Allen}, \& {van der Kruit}}]{Holwerda05c}
{Holwerda}, B.~W., {Gonz\'alez}, R.~A., {Allen}, R.~J., \& {van der Kruit},
  P.~C. 2005{\natexlab{c}}, \aap, 444, 101

\bibitem[{{Holwerda} {et~al.}(2005{\natexlab{d}}){Holwerda}, {Gonz\'alez},
  {Allen}, \& {van der Kruit}}]{Holwerda05e}
{Holwerda}, B.~W., {Gonz\'alez}, R.~A., {Allen}, R.~J., \& {van der Kruit},
  P.~C. 2005{\natexlab{d}}, \aap, 444, 319

\bibitem[{{Holwerda} {et~al.}(2005{\natexlab{e}}){Holwerda}, {Gonz\'alez}, {van
  der Kruit}, \& {Allen}}]{Holwerda05d}
{Holwerda}, B.~W., {Gonz\'alez}, R.~A., {van der Kruit}, P.~C., \& {Allen},
  R.~J. 2005{\natexlab{e}}, \aap, 444, 109

\bibitem[{{Holwerda} {et~al.}(2007{\natexlab{b}}){Holwerda}, {Keel}, \&
  {Bolton}}]{Holwerda07c}
{Holwerda}, B.~W., {Keel}, W.~C., \& {Bolton}, A. 2007{\natexlab{b}}, \aj, 134,
  2385

\bibitem[{{Holwerda} {et~al.}(2009){Holwerda}, {Keel}, {Williams}, {Dalcanton},
  \& {de Jong}}]{Holwerda09}
{Holwerda}, B.~W., {Keel}, W.~C., {Williams}, B., {Dalcanton}, J.~J., \& {de
  Jong}, R.~S. 2009, \aj, 137, 3000

\bibitem[{{Holwerda} {et~al.}(2007{\natexlab{c}}){Holwerda}, {Meyer}, {Regan},
  {Calzetti}, {Gordon}, {Smith}, {Dale}, {Engelbracht}, {Jarrett}, {Thornley},
  {Bot}, {Buckalew}, {Kennicutt}, \& {Gonz{\'a}lez}}]{Holwerda07b}
{Holwerda}, B.~W., {Meyer}, M., {Regan}, M., {et~al.} 2007{\natexlab{c}}, \aj,
  134, 1655

\bibitem[{{Israel}(1997)}]{Israel97}
{Israel}, F.~P. 1997, \aap, 328, 471

\bibitem[{{Kannappan}(2004)}]{Kannappan04}
{Kannappan}, S.~J. 2004, \apjl, 611, L89

\bibitem[{{Keel} {et~al.}(2012, {\em submitted}){Keel}, {Manning}, {Holwerda},
  {Mezzoprete}, {Lintott}, \& {Schawinski}}]{Keel11}
{Keel}, W.~C., {Manning}, A.~M., {Holwerda}, B.~W., {et~al.} 2012, {\em
  submitted}, MNRAS

\bibitem[{{Keel} \& {White}(2001{\natexlab{a}})}]{kw01a}
{Keel}, W.~C. \& {White}, III, R.~E. 2001{\natexlab{a}}, \aj, 121, 1442

\bibitem[{{Keel} \& {White}(2001{\natexlab{b}})}]{kw01b}
{Keel}, W.~C. \& {White}, III, R.~E. 2001{\natexlab{b}}, \aj, 122, 1369

\bibitem[{{Kennicutt} {et~al.}(2003){Kennicutt}, {Armus}, {Bendo}, {Calzetti},
  {Dale}, {Draine}, {Engelbracht}, {Gordon}, {Grauer}, {Helou}, {Hollenbach},
  {Jarrett}, {Kewley}, {Leitherer}, {Li}, {Malhotra}, {Regan}, {Rieke},
  {Rieke}, {Roussel}, {Smith}, {Thornley}, \& {Walter}}]{SINGS}
{Kennicutt}, R.~C., {Armus}, L., {Bendo}, G., {et~al.} 2003, \pasp, 115, 928

\bibitem[{{Kennicutt} {et~al.}(2011){Kennicutt}, {Calzetti}, {Aniano},
  {Appleton}, {Armus}, {Beir{\~a}o}, {Bolatto}, {Brandl}, {Crocker}, {Croxall},
  {Dale}, {Meyer}, {Draine}, {Engelbracht}, {Galametz}, {Gordon}, {Groves},
  {Hao}, {Helou}, {Hinz}, {Hunt}, {Johnson}, {Koda}, {Krause}, {Leroy}, {Li},
  {Meidt}, {Montiel}, {Murphy}, {Rahman}, {Rix}, {Roussel}, {Sandstrom},
  {Sauvage}, {Schinnerer}, {Skibba}, {Smith}, {Srinivasan}, {Vigroux},
  {Walter}, {Wilson}, {Wolfire}, \& {Zibetti}}]{Kennicutt11}
{Kennicutt}, R.~C., {Calzetti}, D., {Aniano}, G., {et~al.} 2011, \pasp, 123,
  1347

\bibitem[{{Kennicutt}(1998)}]{Kennicutt98}
{Kennicutt}, Jr., R.~C. 1998, \apj, 498, 541

\bibitem[{{Kennicutt} {et~al.}(2007){Kennicutt}, {Calzetti}, {Walter}, {Helou},
  {Hollenbach}, {Armus}, {Bendo}, {Dale}, {Draine}, {Engelbracht}, {Gordon},
  {Prescott}, {Regan}, {Thornley}, {Bot}, {Brinks}, {de Blok}, {de Mello},
  {Meyer}, {Moustakas}, {Murphy}, {Sheth}, \& {Smith}}]{Kennicutt07}
{Kennicutt}, Jr., R.~C., {Calzetti}, D., {Walter}, F., {et~al.} 2007, \apj,
  671, 333

\bibitem[{{Kobulnicky} \& {Kewley}(2004)}]{KK04}
{Kobulnicky}, H.~A. \& {Kewley}, L.~J. 2004, \apj, 617, 240

\bibitem[{{Leroy} {et~al.}(2007){Leroy}, {Bolatto}, {Stanimirovic}, {Mizuno},
  {Israel}, \& {Bot}}]{Leroy07}
{Leroy}, A., {Bolatto}, A., {Stanimirovic}, S., {et~al.} 2007, \apj, 658, 1027

\bibitem[{{Leroy} {et~al.}(2011){Leroy}, {Bolatto}, {Gordon}, {Sandstrom},
  {Gratier}, {Rosolowsky}, {Engelbracht}, {Mizuno}, {Corbelli}, {Fukui}, \&
  {Kawamura}}]{Leroy11}
{Leroy}, A.~K., {Bolatto}, A., {Gordon}, K., {et~al.} 2011, ArXiv
  e-prints/1102.4618

\bibitem[{{Leroy} {et~al.}(2009){Leroy}, {Walter}, {Bigiel}, {Usero}, {Weiss},
  {Brinks}, {de Blok}, {Kennicutt}, {Schuster}, {Kramer}, {Wiesemeyer}, \&
  {Roussel}}]{heracles}
{Leroy}, A.~K., {Walter}, F., {Bigiel}, F., {et~al.} 2009, \aj, 137, 4670

\bibitem[{{Leroy} {et~al.}(2008){Leroy}, {Walter}, {Brinks}, {Bigiel}, {de
  Blok}, {Madore}, \& {Thornley}}]{Leroy08}
{Leroy}, A.~K., {Walter}, F., {Brinks}, E., {et~al.} 2008, \aj, 136, 2782

\bibitem[{{Li} \& {Draine}(2001)}]{Draine01b}
{Li}, A. \& {Draine}, B.~T. 2001, \apj, 554, 778

\bibitem[{{Lintott} {et~al.}(2008){Lintott}, {Schawinski}, {Slosar}, {Land},
  {Bamford}, {Thomas}, {Raddick}, {Nichol}, {Szalay}, {Andreescu}, {Murray}, \&
  {Vandenberg}}]{Lintott08}
{Lintott}, C.~J., {Schawinski}, K., {Slosar}, A., {et~al.} 2008, \mnras, 389,
  1179

\bibitem[{{Mac Low} \& {Glover}(2012)}]{Mac-Low12}
{Mac Low}, M.-M. \& {Glover}, S.~C.~O. 2012, \apj, 746, 135

\bibitem[{{MacGillivray}(1975)}]{McGillivray75}
{MacGillivray}, H.~T. 1975, \mnras, 170, 241

\bibitem[{{Madden} {et~al.}(2011){Madden}, {Galametz}, {Cormier},
  {Lebouteiller}, {Galliano}, {Hony}, {Remy}, {Sauvage}, {Contursi}, {Sturm},
  {Poglitsch}, {Pohlen}, {Smith}, {Bendo}, {O'Halloran}, {SPIRE SAG 2}, \&
  {consortia.}}]{Madden11}
{Madden}, S.~C., {Galametz}, M., {Cormier}, D., {et~al.} 2011, ArXiv e-prints

\bibitem[{{Madden} {et~al.}(1997){Madden}, {Poglitsch}, {Geis}, {Stacey}, \&
  {Townes}}]{Madden97}
{Madden}, S.~C., {Poglitsch}, A., {Geis}, N., {Stacey}, G.~J., \& {Townes},
  C.~H. 1997, \apj, 483, 200

\bibitem[{{Magrini} {et~al.}(2011){Magrini}, {Bianchi}, {Corbelli}, {Cortese},
  {Hunt}, {Smith}, {Vlahakis}, {Davies}, {Bendo}, {Baes}, {Boselli}, {Clemens},
  {Casasola}, {De Looze}, {Fritz}, {Giovanardi}, {Grossi}, {Hughes}, {Madden},
  {Pappalardo}, {Pohlen}, {di Serego Alighieri}, \& {Verstappen}}]{Magrini11}
{Magrini}, L., {Bianchi}, S., {Corbelli}, E., {et~al.} 2011, ArXiv
  e-prints/1106.0618

\bibitem[{{Moustakas} {et~al.}(2010){Moustakas}, {Kennicutt}, {Tremonti},
  {Dale}, {Smith}, \& {Calzetti}}]{Moustakas10}
{Moustakas}, J., {Kennicutt}, Jr., R.~C., {Tremonti}, C.~A., {et~al.} 2010,
  \apjs, 190, 233

\bibitem[{{Mu{\~n}oz-Mateos} {et~al.}(2011){Mu{\~n}oz-Mateos}, {Boissier}, {Gil
  de Paz}, {Zamorano}, {Kennicutt}, {Moustakas}, {Prantzos}, \&
  {Gallego}}]{Munoz-Mateos11}
{Mu{\~n}oz-Mateos}, J.~C., {Boissier}, S., {Gil de Paz}, A., {et~al.} 2011,
  \apj, 731, 10

\bibitem[{{Mu{\~n}oz-Mateos} {et~al.}(2009{\natexlab{a}}){Mu{\~n}oz-Mateos},
  {Gil de Paz}, {Boissier}, {Zamorano}, {Dale}, {P{\'e}rez-Gonz{\'a}lez},
  {Gallego}, {Madore}, {Bendo}, {Thornley}, {Draine}, {Boselli}, {Buat},
  {Calzetti}, {Moustakas}, \& {Kennicutt}}]{Munoz-Mateos09b}
{Mu{\~n}oz-Mateos}, J.~C., {Gil de Paz}, A., {Boissier}, S., {et~al.}
  2009{\natexlab{a}}, \apj, 701, 1965

\bibitem[{{Mu{\~n}oz-Mateos} {et~al.}(2009{\natexlab{b}}){Mu{\~n}oz-Mateos},
  {Gil de Paz}, {Zamorano}, {Boissier}, {Dale}, {P{\'e}rez-Gonz{\'a}lez},
  {Gallego}, {Madore}, {Bendo}, {Boselli}, {Buat}, {Calzetti}, {Moustakas}, \&
  {Kennicutt}}]{Munoz-Mateos09a}
{Mu{\~n}oz-Mateos}, J.~C., {Gil de Paz}, A., {Zamorano}, J., {et~al.}
  2009{\natexlab{b}}, \apj, 703, 1569

\bibitem[{{Narayanan}(2011)}]{Narayanan11b}
{Narayanan}, D. 2011, ArXiv e-prints

\bibitem[{{Narayanan} {et~al.}(2011){Narayanan}, {Krumholz}, {Ostriker}, \&
  {Hernquist}}]{Narayanan11c}
{Narayanan}, D., {Krumholz}, M., {Ostriker}, E.~C., \& {Hernquist}, L. 2011,
  \mnras, 418, 664

\bibitem[{{Narayanan} {et~al.}(2012){Narayanan}, {Krumholz}, {Ostriker}, \&
  {Hernquist}}]{Narayanan12}
{Narayanan}, D., {Krumholz}, M.~R., {Ostriker}, E.~C., \& {Hernquist}, L. 2012,
  \mnras, 2537

\bibitem[{{Obreschkow} {et~al.}(2009){Obreschkow}, {Croton}, {DeLucia},
  {Khochfar}, \& {Rawlings}}]{Obreschkow09c}
{Obreschkow}, D., {Croton}, D., {DeLucia}, G., {Khochfar}, S., \& {Rawlings},
  S. 2009, \apj, 698, 1467

\bibitem[{{Oosterloo} {et~al.}(2007){Oosterloo}, {Fraternali}, \&
  {Sancisi}}]{Oosterloo07}
{Oosterloo}, T., {Fraternali}, F., \& {Sancisi}, R. 2007, \aj, 134, 1019

\bibitem[{{Pflamm-Altenburg} \& {Kroupa}(2008)}]{Pflamm-Altenburg08}
{Pflamm-Altenburg}, J. \& {Kroupa}, P. 2008, \nat, 455, 641

\bibitem[{{Pilyugin} \& {Thuan}(2005)}]{PT05}
{Pilyugin}, L.~S. \& {Thuan}, T.~X. 2005, \apj, 631, 231

\bibitem[{{Planck Collaboration} {et~al.}(2011){Planck Collaboration}, {Ade},
  {Aghanim}, {Arnaud}, {Ashdown}, {Aumont}, {Baccigalupi}, {Balbi}, {Banday},
  {Barreiro}, \& et~al.}]{Planck-Collaboration11a}
{Planck Collaboration}, {Ade}, P.~A.~R., {Aghanim}, N., {et~al.} 2011, \aap,
  536, A19

\bibitem[{{Pohlen} {et~al.}(2010){Pohlen}, {Cortese}, {Smith}, {Eales},
  {Boselli}, {Bendo}, {Gomez}, {Papageorgiou}, {Auld}, {Baes}, {Bock},
  {Bradford}, {Buat}, {Castro-Rodriguez}, {Chanial}, {Charlot}, {Ciesla},
  {Clements}, {Cooray}, {Cormier}, {Dwek}, {Eales}, {Elbaz}, {Galametz},
  {Galliano}, {Gear}, {Glenn}, {Griffin}, {Hony}, {Isaak}, {Levenson}, {Lu},
  {Madden}, {O'Halloran}, {Okumura}, {Oliver}, {Page}, {Panuzzo}, {Parkin},
  {Perez-Fournon}, {Rangwala}, {Rigby}, {Roussel}, {Rykala}, {Sacchi},
  {Sauvage}, {Schulz}, {Schirm}, {Smith}, {Spinoglio}, {Stevens}, {Srinivasan},
  {Symeonidis}, {Trichas}, {Vaccari}, {Vigroux}, {Wilson}, {Wozniak}, {Wright},
  \& {Zeiliner}}]{Pohlen10}
{Pohlen}, M., {Cortese}, L., {Smith}, M.~W.~L., {et~al.} 2010, ArXiv e-prints

\bibitem[{{Popescu} {et~al.}(2000){Popescu}, {Misiriotis}, {Kylafis}, {Tuffs},
  \& {Fischera}}]{Popescu00}
{Popescu}, C.~C., {Misiriotis}, A., {Kylafis}, N.~D., {Tuffs}, R.~J., \&
  {Fischera}, J. 2000, \aap, 362, 138

\bibitem[{{Popescu} \& {Tuffs}(2002)}]{Popescu02}
{Popescu}, C.~C. \& {Tuffs}, R.~J. 2002, Reviews of Modern Astronomy, 15, 239

\bibitem[{{Roman-Duval} {et~al.}(2010){Roman-Duval}, {Israel}, {Bolatto},
  {Hughes}, {Leroy}, {Meixner}, {Gordon}, {Madden}, {Paradis}, {Kawamura},
  {Li}, {Sauvage}, {Wong}, {Bernard}, {Engelbracht}, {Hony}, {Kim}, {Misselt},
  {Okumura}, {Ott}, {Panuzzo}, {Pineda}, {Reach}, \& {Rubio}}]{Roman-Duval10}
{Roman-Duval}, J., {Israel}, F.~P., {Bolatto}, A., {et~al.} 2010, \aap, 518,
  L74+

\bibitem[{{Rosolowsky}(2005)}]{Rosolowsky05a}
{Rosolowsky}, E. 2005, \pasp, 117, 1403

\bibitem[{{Sancisi} {et~al.}(2008){Sancisi}, {Fraternali}, {Oosterloo}, \& {van
  der Hulst}}]{Sancisi08}
{Sancisi}, R., {Fraternali}, F., {Oosterloo}, T., \& {van der Hulst}, T. 2008,
  \aapr, 15, 189

\bibitem[{{Sandstrom} {et~al.}(2011){Sandstrom}, {Leroy}, {Walter}, {Aniano},
  {Draine}, {Calzetti}, {Kennicutt}, \& {KINGFISH Team}}]{Sandstrom11}
{Sandstrom}, K.~M., {Leroy}, A.~K., {Walter}, F., {et~al.} 2011, in Bulletin of
  the American Astronomical Society, Vol.~43, American Astronomical Society
  Meeting Abstracts \#217, \#202.07--+

\bibitem[{{Schruba} {et~al.}(2011){Schruba}, {Leroy}, {Walter}, {Bigiel},
  {Brinks}, {de Blok}, {Dumas}, {Kramer}, {Rosolowsky}, {Sandstrom},
  {Schuster}, {Usero}, {Weiss}, \& {Wiesemeyer}}]{Schruba11}
{Schruba}, A., {Leroy}, A.~K., {Walter}, F., {et~al.} 2011, ArXiv e-prints

\bibitem[{{Shapley}(1951)}]{Shapley51}
{Shapley}, H. 1951, Proceedings of the National Academy of Science, 37, 133

\bibitem[{{Shetty} {et~al.}(2011){Shetty}, {Glover}, {Dullemond}, {Ostriker},
  {Harris}, \& {Klessen}}]{Shetty11}
{Shetty}, R., {Glover}, S.~C., {Dullemond}, C.~P., {et~al.} 2011, ArXiv
  e-prints

\bibitem[{{Skibba} {et~al.}(2011){Skibba}, {Engelbracht}, {Dale}, {Hinz},
  {Zibetti}, {Crocker}, {Groves}, {Hunt}, {Johnson}, {Meidt}, {Murphy},
  {Appleton}, {Armus}, {Bolatto}, {Brandl}, {Calzetti}, {Croxall}, {Galametz},
  {Gordon}, {Kennicutt}, {Koda}, {Krause}, {Montiel}, {Rix}, {Roussel},
  {Sandstrom}, {Sauvage}, {Schinnerer}, {Smith}, {Walter}, {Wilson}, \&
  {Wolfire}}]{Skibba11a}
{Skibba}, R.~A., {Engelbracht}, C.~W., {Dale}, D., {et~al.} 2011, ArXiv
  e-prints

\bibitem[{{Smith} {et~al.}(2000){Smith}, {Allen}, {Bohlin}, {Nicholson}, \&
  {Stecher}}]{Smith00}
{Smith}, D.~A., {Allen}, R.~J., {Bohlin}, R.~C., {Nicholson}, N., \& {Stecher},
  T.~P. 2000, \apj, 538, 608

\bibitem[{{Smith} {et~al.}(2010){Smith}, {Vlahakis}, {Baes}, {Bendo},
  {Bianchi}, {Bomans}, {Boselli}, {Clemens}, {Corbelli}, {Cortese}, {Dariush},
  {Davies}, {de Looze}, {di Serego Alighieri}, {Fadda}, {Fritz},
  {Garcia-Appadoo}, {Gavazzi}, {Giovanardi}, {Grossi}, {Hughes}, {Hunt},
  {Jones}, {Madden}, {Pierini}, {Pohlen}, {Sabatini}, {Verstappen}, {Xilouris},
  \& {Zibetti}}]{Smith10b}
{Smith}, M.~W.~L., {Vlahakis}, C., {Baes}, M., {et~al.} 2010, \aap, 518, L51+

\bibitem[{{Thilker} {et~al.}(2007){Thilker}, {Boissier}, {Bianchi}, {Calzetti},
  {Boselli}, {Dale}, {Seibert}, {Braun}, {Burgarella}, {Gil de Paz}, {Helou},
  {Walter}, {Kennicutt}, {Madore}, {Martin}, {Barlow}, {Forster}, {Friedman},
  {Morrissey}, {Neff}, {Schiminovich}, {Small}, {Wyder}, {Donas}, {Heckman},
  {Lee}, {Milliard}, {Rich}, {Szalay}, {Welsh}, \& {Yi}}]{Thilker07a}
{Thilker}, D.~A., {Boissier}, S., {Bianchi}, L., {et~al.} 2007, \apjs, 173, 572

\bibitem[{{Walter} {et~al.}(2008){Walter}, {Brinks}, {de Blok}, {Bigiel},
  {Kennicutt}, {Thornley}, \& {Leroy}}]{Walter08}
{Walter}, F., {Brinks}, E., {de Blok}, W.~J.~G., {et~al.} 2008, \aj, 136, 2563

\bibitem[{{Walter} {et~al.}(2009){Walter}, {Leroy}, {Bigiel}, {Schruba},
  {Usero}, {Kennicutt}, \& {HERACLES team}}]{Walter09}
{Walter}, F., {Leroy}, A., {Bigiel}, F., {et~al.} 2009, in American
  Astronomical Society Meeting Abstracts, Vol. 214, American Astronomical
  Society Meeting Abstracts, \#419.08--+

\bibitem[{{Weingartner} \& {Draine}(2001)}]{Weingartner01b}
{Weingartner}, J.~C. \& {Draine}, B.~T. 2001, \apj, 553, 581

\bibitem[{{Wesselink}(1961)}]{Wesselink61b}
{Wesselink}, A.~J. 1961, \mnras, 122, 509

\bibitem[{{West} {et~al.}(2010){West}, {Garcia-Appadoo}, {Dalcanton}, {Disney},
  {Rockosi}, {Ivezi{\'c}}, {Bentz}, \& {Brinkmann}}]{West10a}
{West}, A.~A., {Garcia-Appadoo}, D.~A., {Dalcanton}, J.~J., {et~al.} 2010, \aj,
  139, 315

\bibitem[{{White} {et~al.}(2000){White}, {Keel}, \& {Conselice}}]{kw00a}
{White}, III, R.~E., {Keel}, W.~C., \& {Conselice}, C.~J. 2000, \apj, 542, 761

\bibitem[{{Wolfire} {et~al.}(2010){Wolfire}, {Hollenbach}, \&
  {McKee}}]{Wolfire10}
{Wolfire}, M.~G., {Hollenbach}, D., \& {McKee}, C.~F. 2010, \apj, 716, 1191

\bibitem[{{Zaritsky}(1994)}]{Zaritsky94}
{Zaritsky}, D. 1994, \aj, 108, 1619

\end{thebibliography}

\newpage

\appendix

%\begin{figure*}
%\begin{center}
%\section*{SED of each WFPC2 field}
%		\includegraphics[width=0.32\textwidth]{../../plot/NGC925.pdf}
%		\includegraphics[width=0.32\textwidth]{../../plot/NGC2841.pdf}
%		\includegraphics[width=0.32\textwidth]{../../plot/NGC3031.pdf}\\
%		\includegraphics[width=0.32\textwidth]{../../plot/NGC3198.pdf}
%		\includegraphics[width=0.32\textwidth]{../../plot/NGC3351.pdf}
%		\includegraphics[width=0.32\textwidth]{../../plot/NGC3621-1.pdf}\\
%		\includegraphics[width=0.32\textwidth]{../../plot/NGC3621-2.pdf}
%		\includegraphics[width=0.32\textwidth]{../../plot/NGC3627.pdf}
%		\includegraphics[width=0.32\textwidth]{../../plot/NGC5194-1.pdf}\\
%		\includegraphics[width=0.32\textwidth]{../../plot/NGC5194-2.pdf}
%		\includegraphics[width=0.32\textwidth]{../../plot/NGC6946.pdf}
%		\includegraphics[width=0.32\textwidth]{../../plot/NGC7331.pdf}
%\caption{The Spectral Energy Distribution from Spitzer IRAC and MIPS \citep{Holwerda07a} and the SPIRE from this paper.}
%\end{center}
%
%\label{f:seds}
%\end{figure*}
%

\section{MAGPHYS SED Model}
\label{s:magphys}

\begin{figure}[h]
   \centering
    \includegraphics[width=0.5\textwidth]{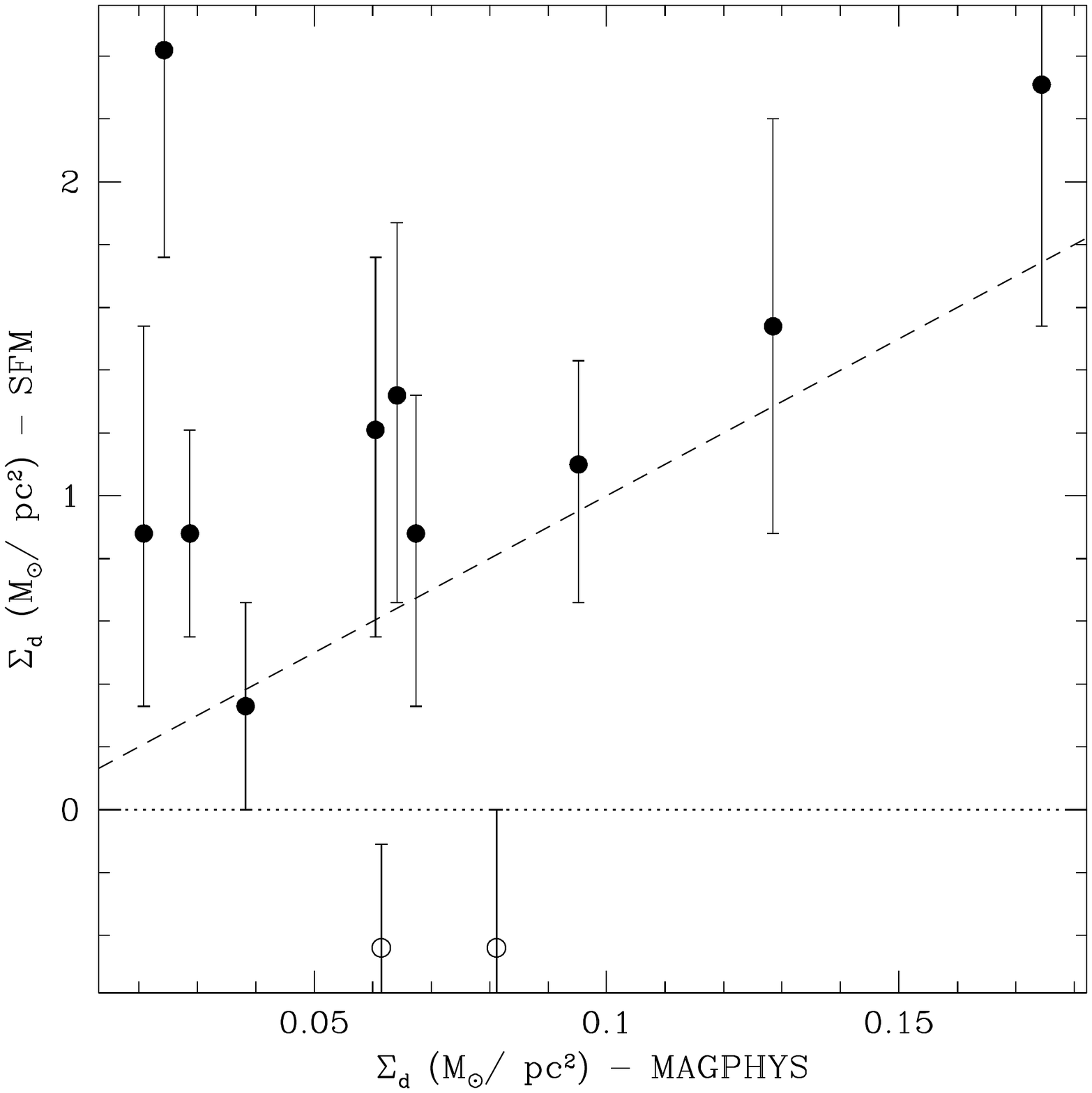}
         \caption{The dust surface densities from the {\sc magphys} fit and inferred from the number of identified background galaxies (SFM) for each WFPC2 aperture. The dashed line denotes a factor ten ratio.
         {\sc magphys} SED models do not take internal structure and differential stellar heating into account.}
     \label{f:magphys}
\end{figure}

As an alternative check of the inferred dust masses, we ran the Multi-wavelength Analysis of Galaxy Physical Properties ({\sc magphys}) package on the {\em Spitzer} and {\em Herschel/SPIRE} surface brightnesses.
This is a self-contained, user-friendly model package to interpret observed spectral energy distributions of galaxies in terms of galaxy-wide physical parameters pertaining to the stars and the interstellar medium, following the approach described in \cite{da-Cunha08}. Figure \ref{f:magphys} summarizes the result: dust surface density derived from the {\sc magphys} fit compared to those inferred from the number of distant galaxies. 
In \cite{Holwerda07a}, we found that the \cite{Draine07} model inferred similar dust optical depths for these disks as the SFM as well as similar (to within a factor two) dust masses.
The discrepancy with {\sc magphys} illustrates, in our view, the importance of modeling sections of spiral disks in resolved observations with more physical models that include a range of stellar heating parameters \cite[e.g. the models by][]{Draine07, Galliano11}.

%Similar to the more direct comparison 
%of Herschel surface brightnesses, the SED fit points to a discrepancy between SED models and SFM derivations \citep[][reached a similar %conclusion]{Holwerda07a}. 
%The relation is mostly linear but offset, which could be accounted for due to dust and star geometry. 

\begin{figure*}
\begin{center}
\section*{SED of each WFPC2 field with the {\sc magphys} fit.}
		\includegraphics[width=0.32\textwidth]{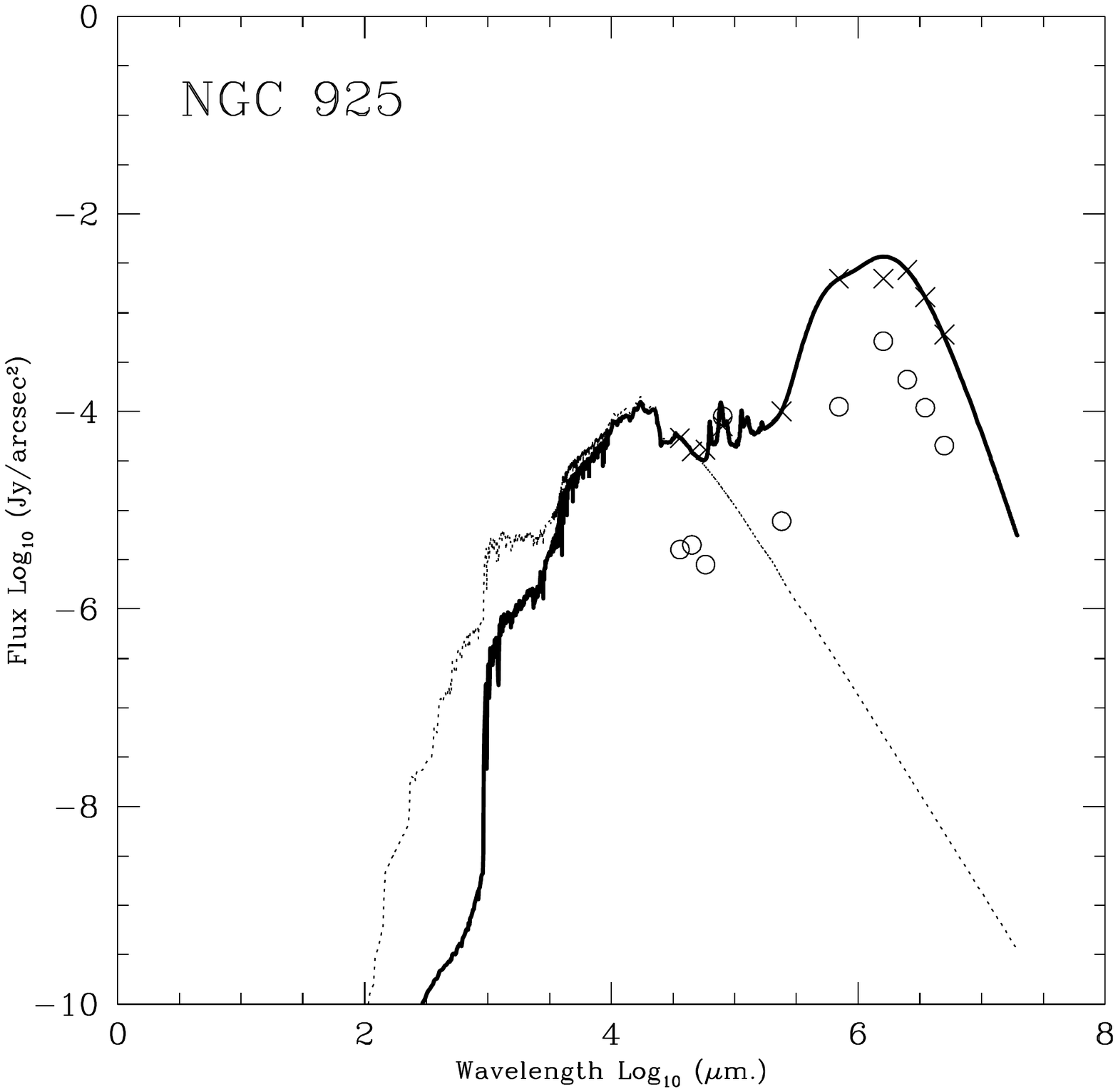}
		\includegraphics[width=0.32\textwidth]{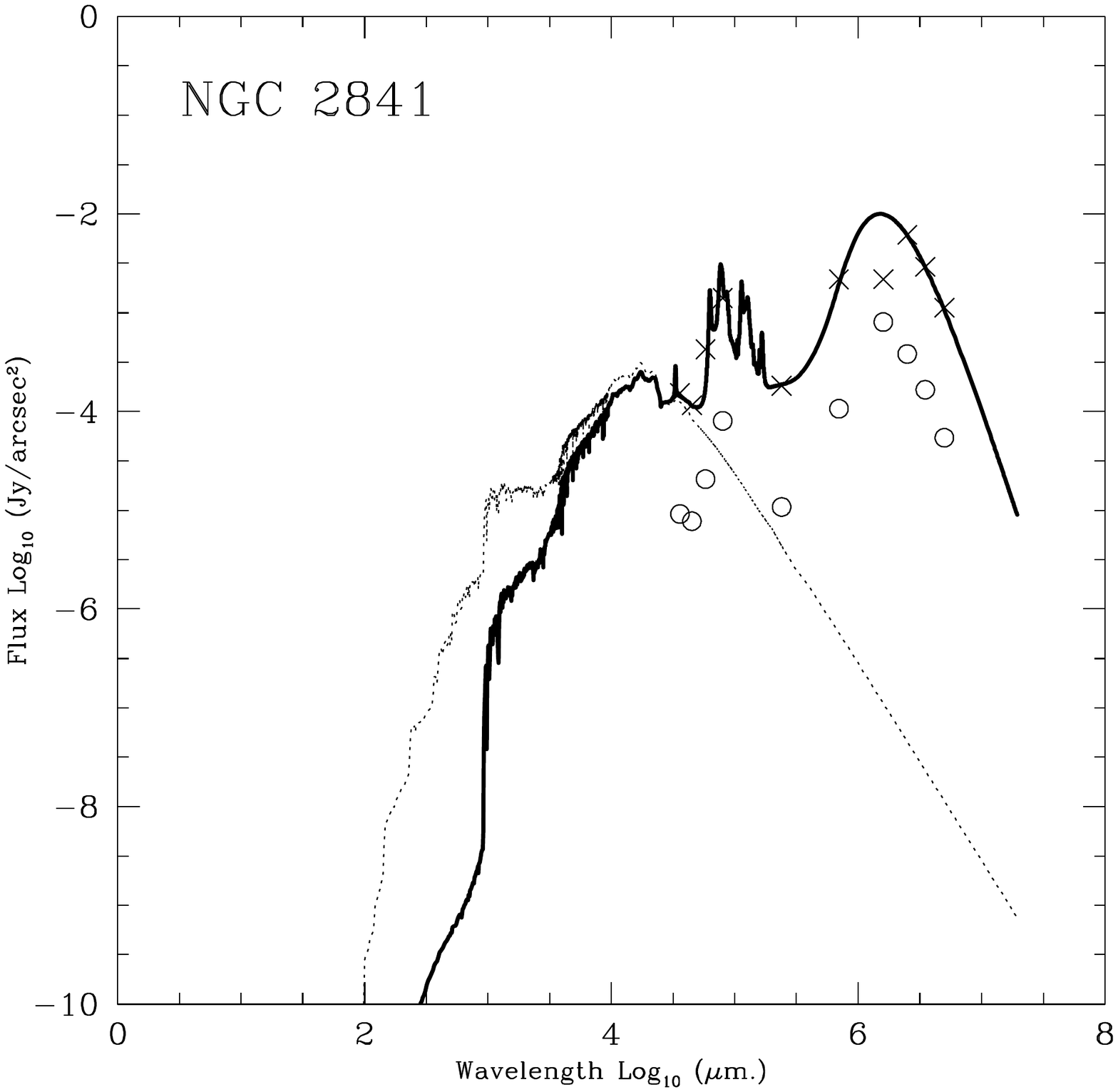}
		\includegraphics[width=0.32\textwidth]{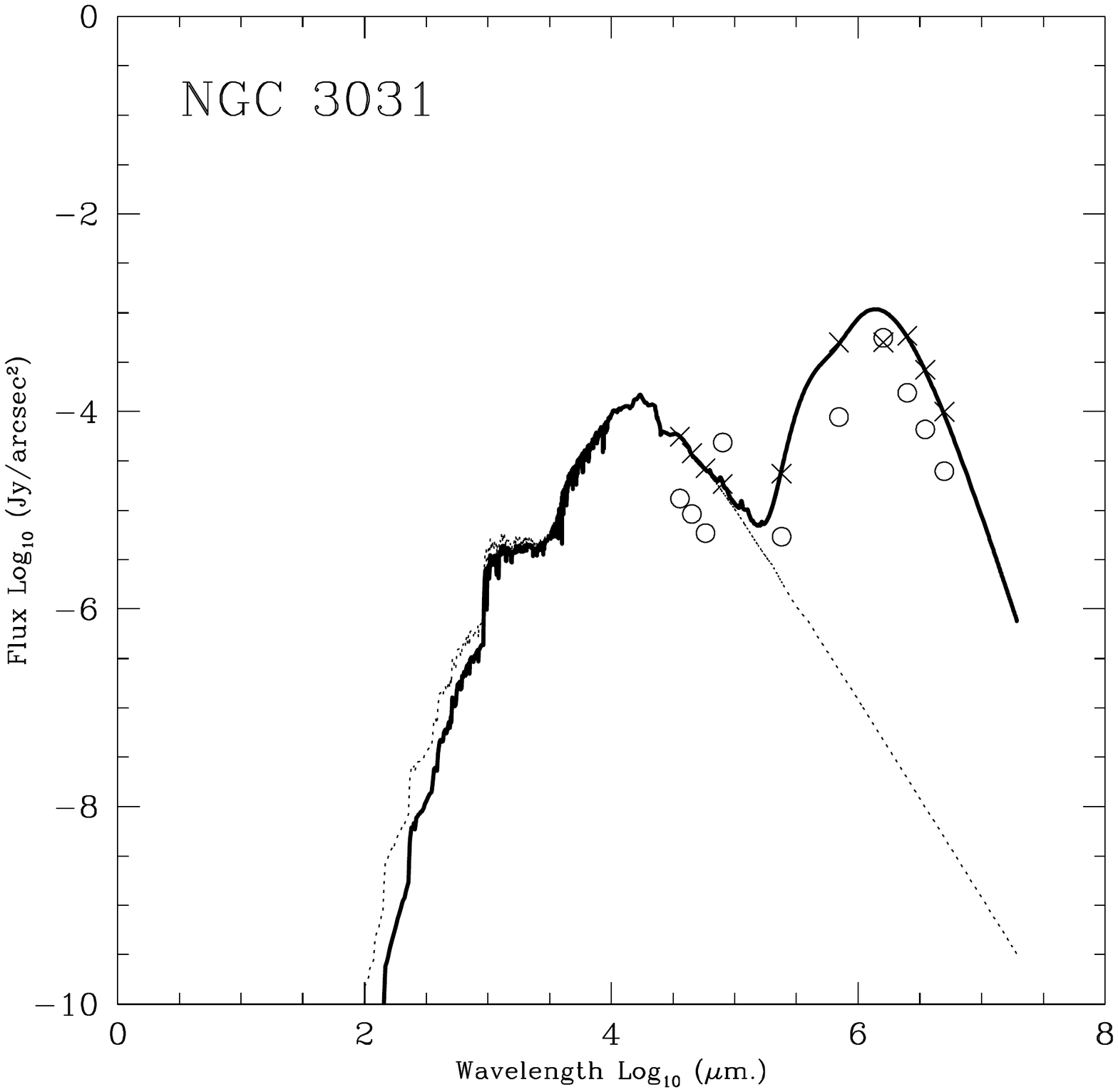}\\
		\includegraphics[width=0.32\textwidth]{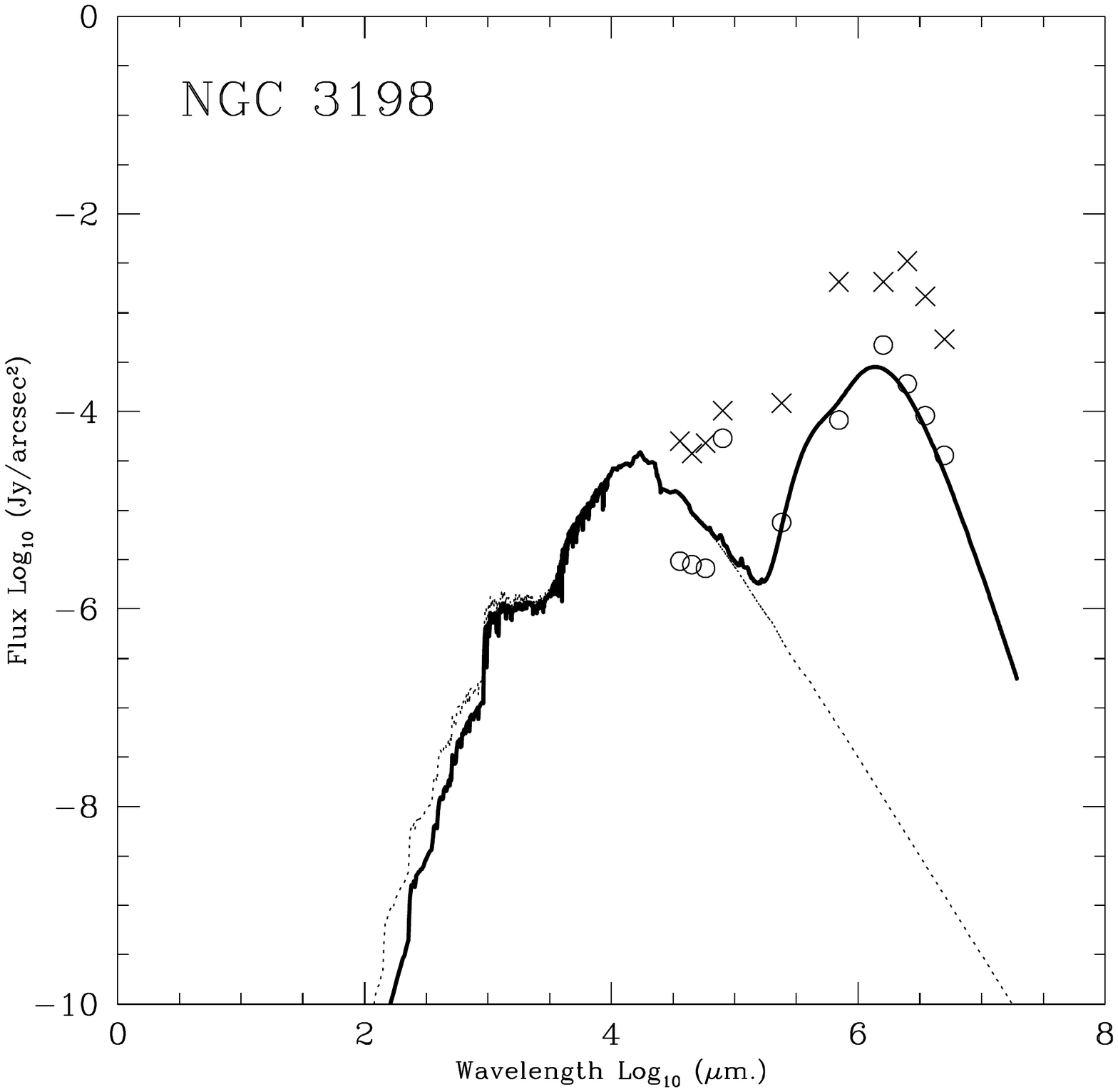}
		\includegraphics[width=0.32\textwidth]{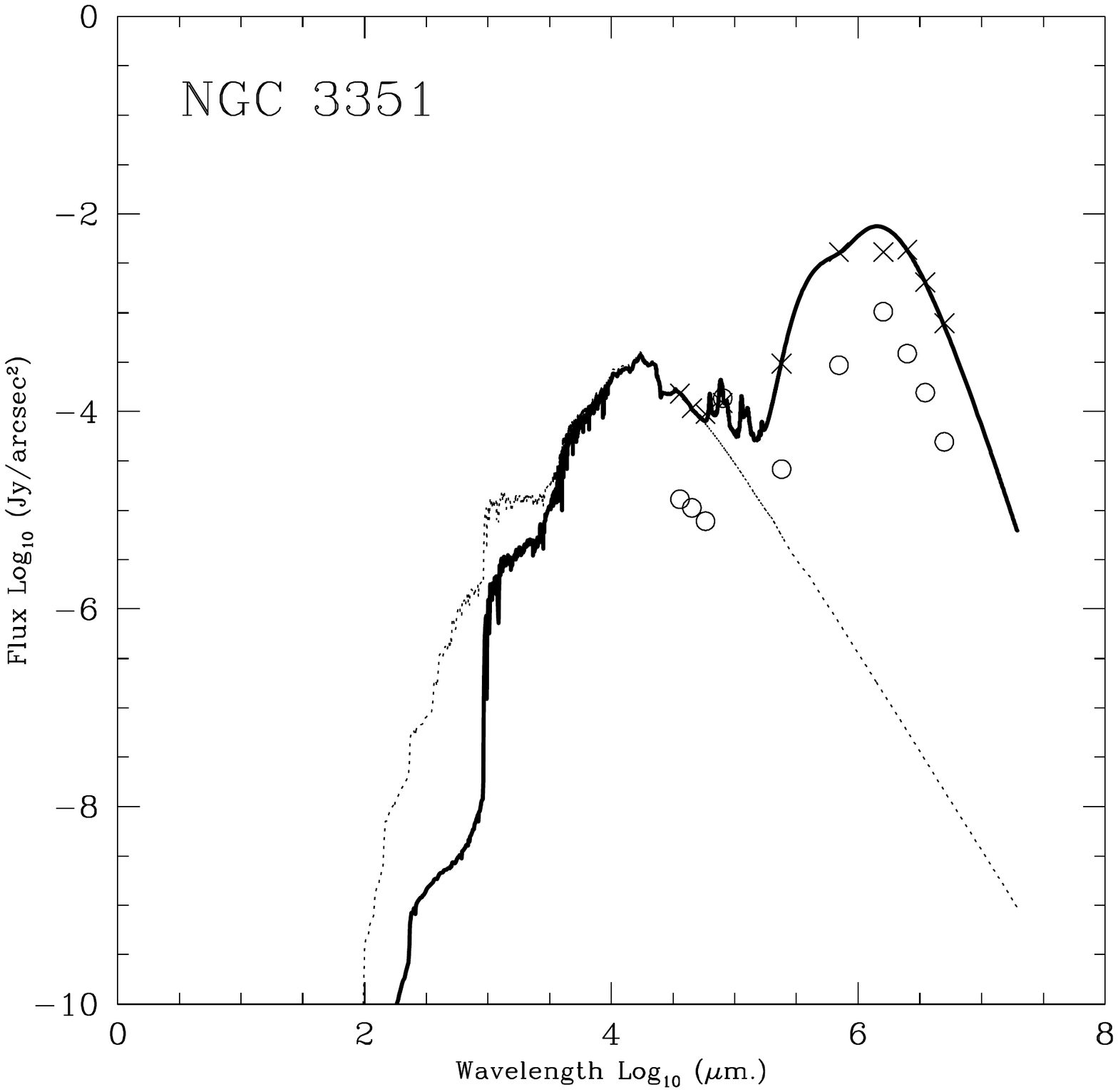}
		\includegraphics[width=0.32\textwidth]{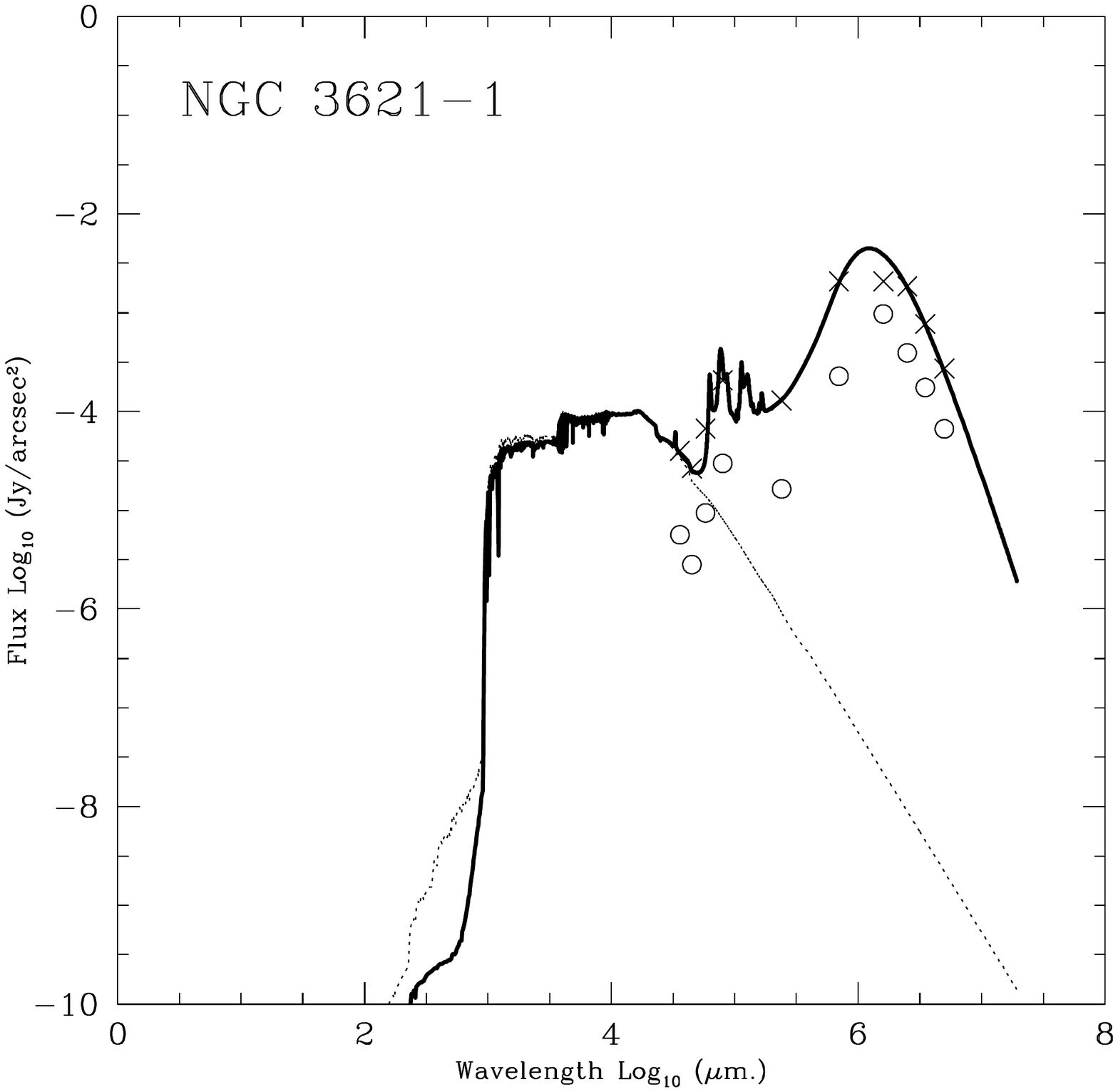}\\
		\includegraphics[width=0.32\textwidth]{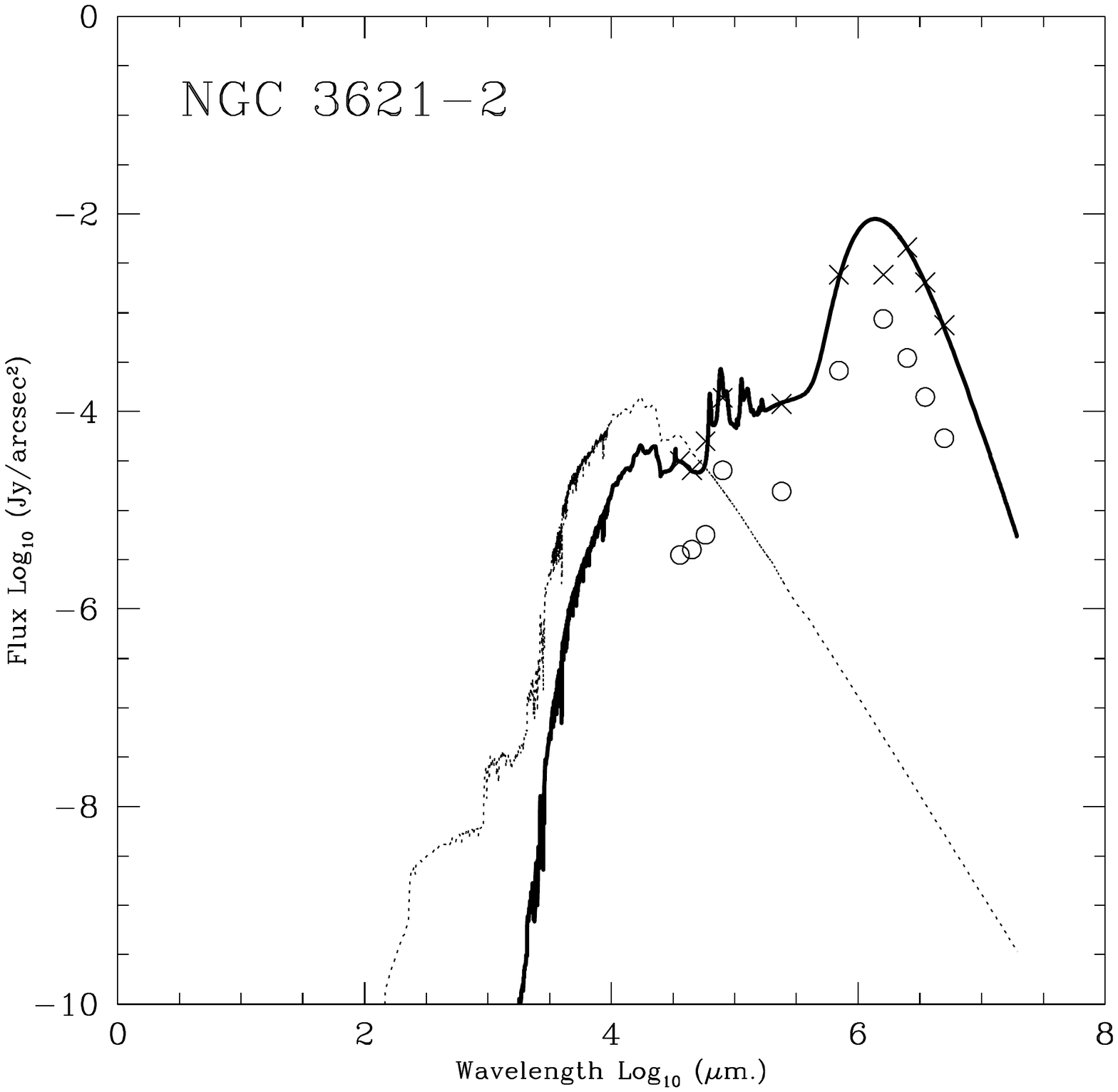}
		\includegraphics[width=0.32\textwidth]{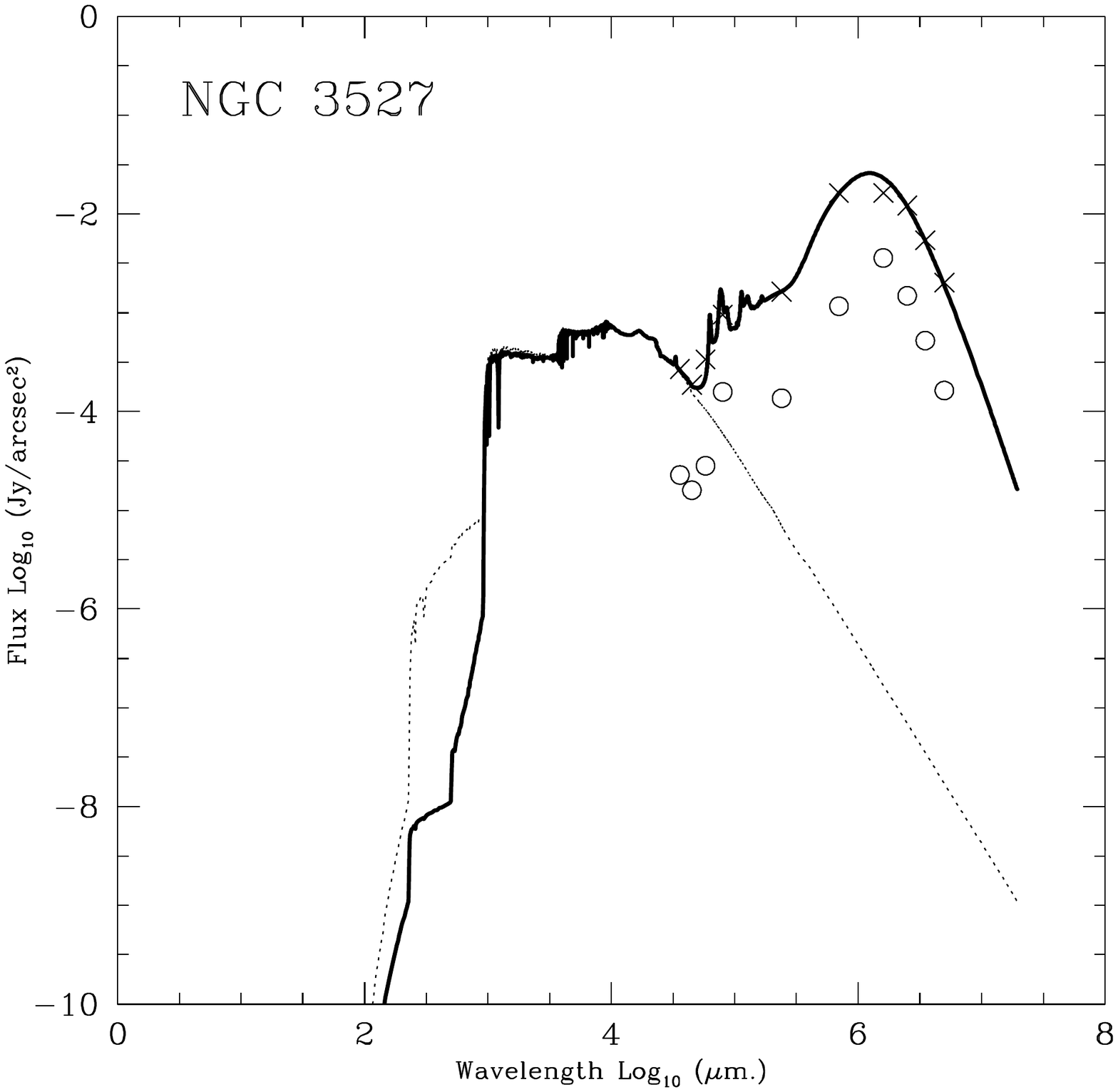}
		\includegraphics[width=0.32\textwidth]{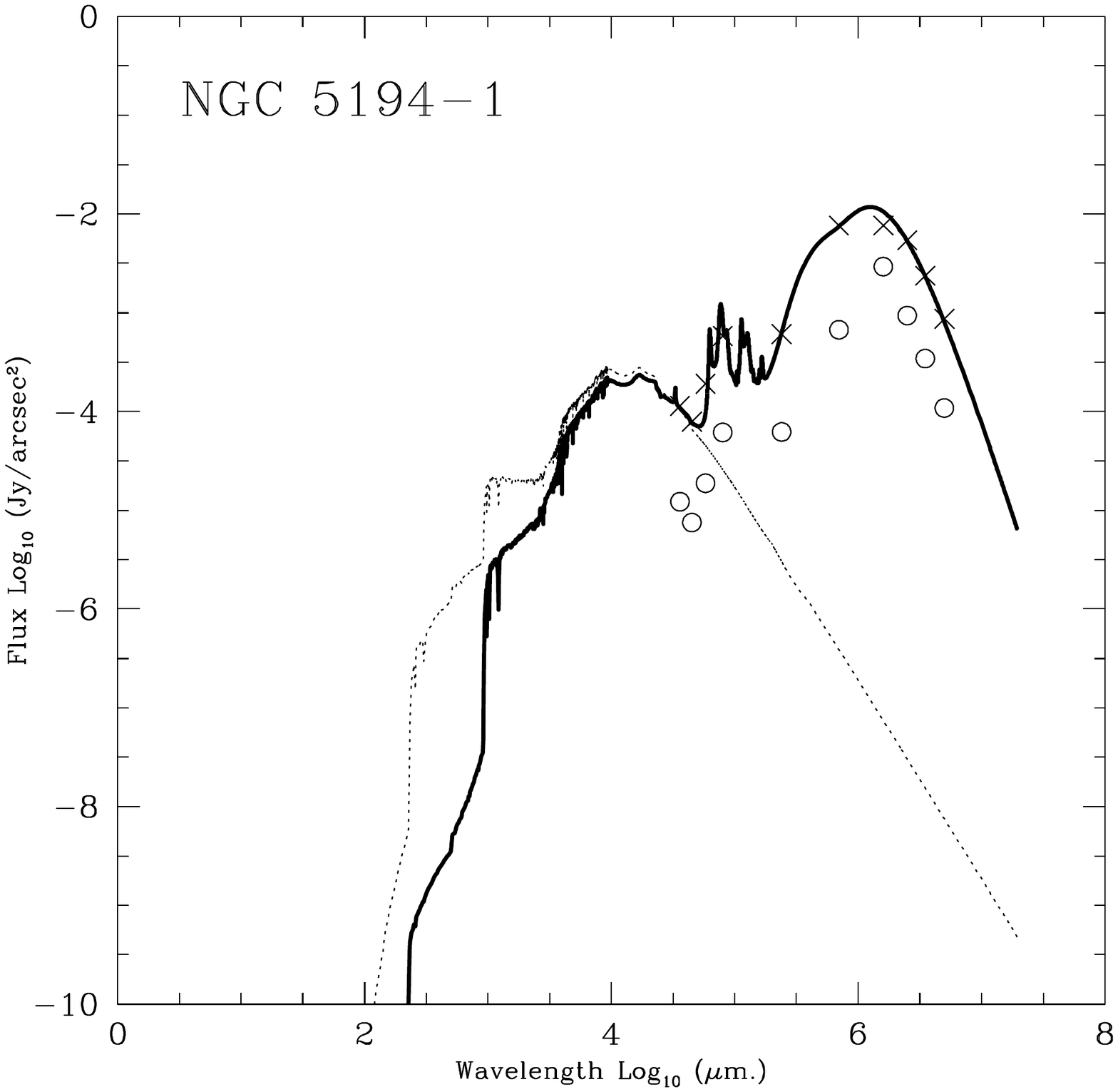}\\
		\includegraphics[width=0.32\textwidth]{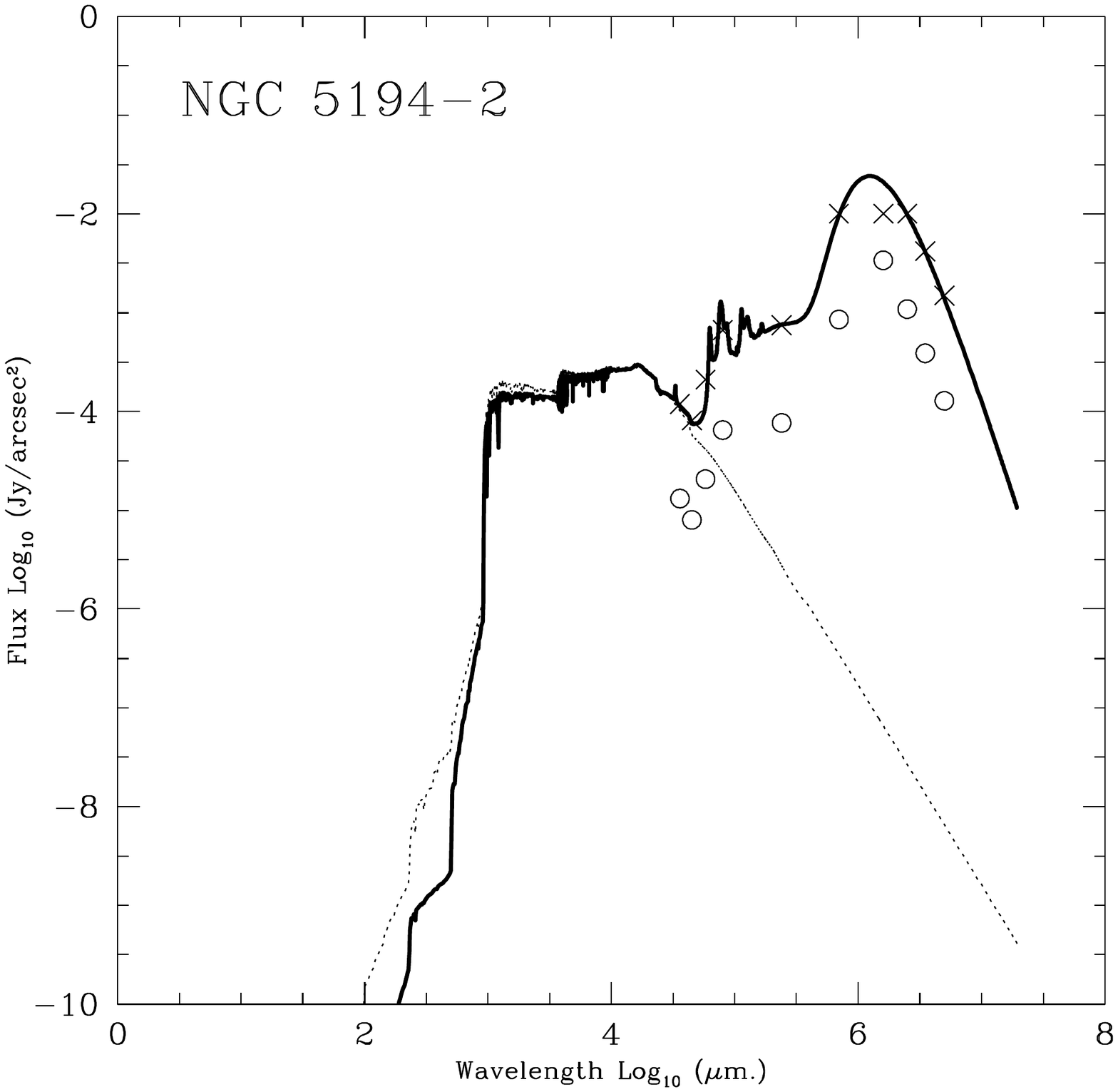}
		\includegraphics[width=0.32\textwidth]{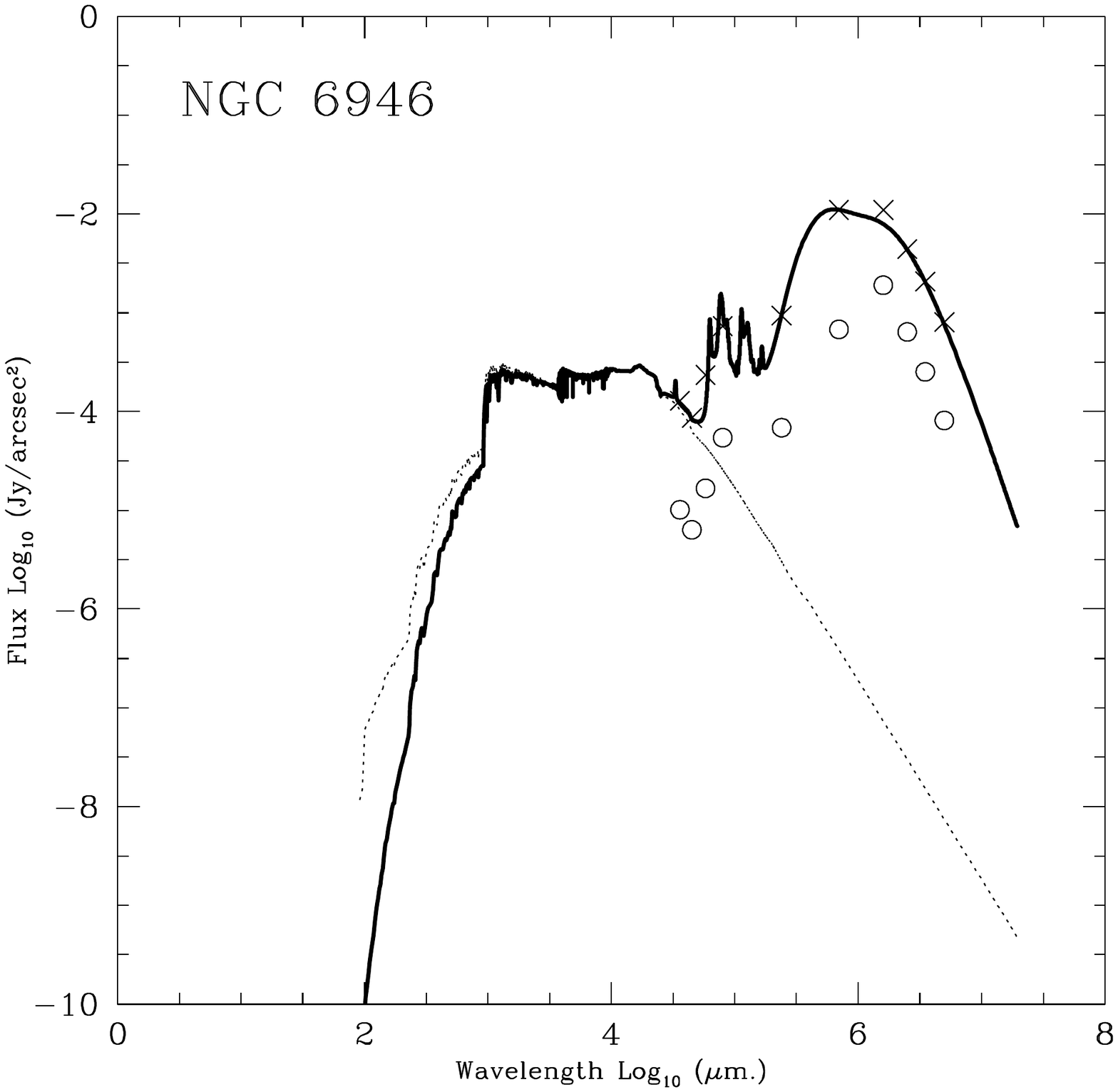}
		\includegraphics[width=0.32\textwidth]{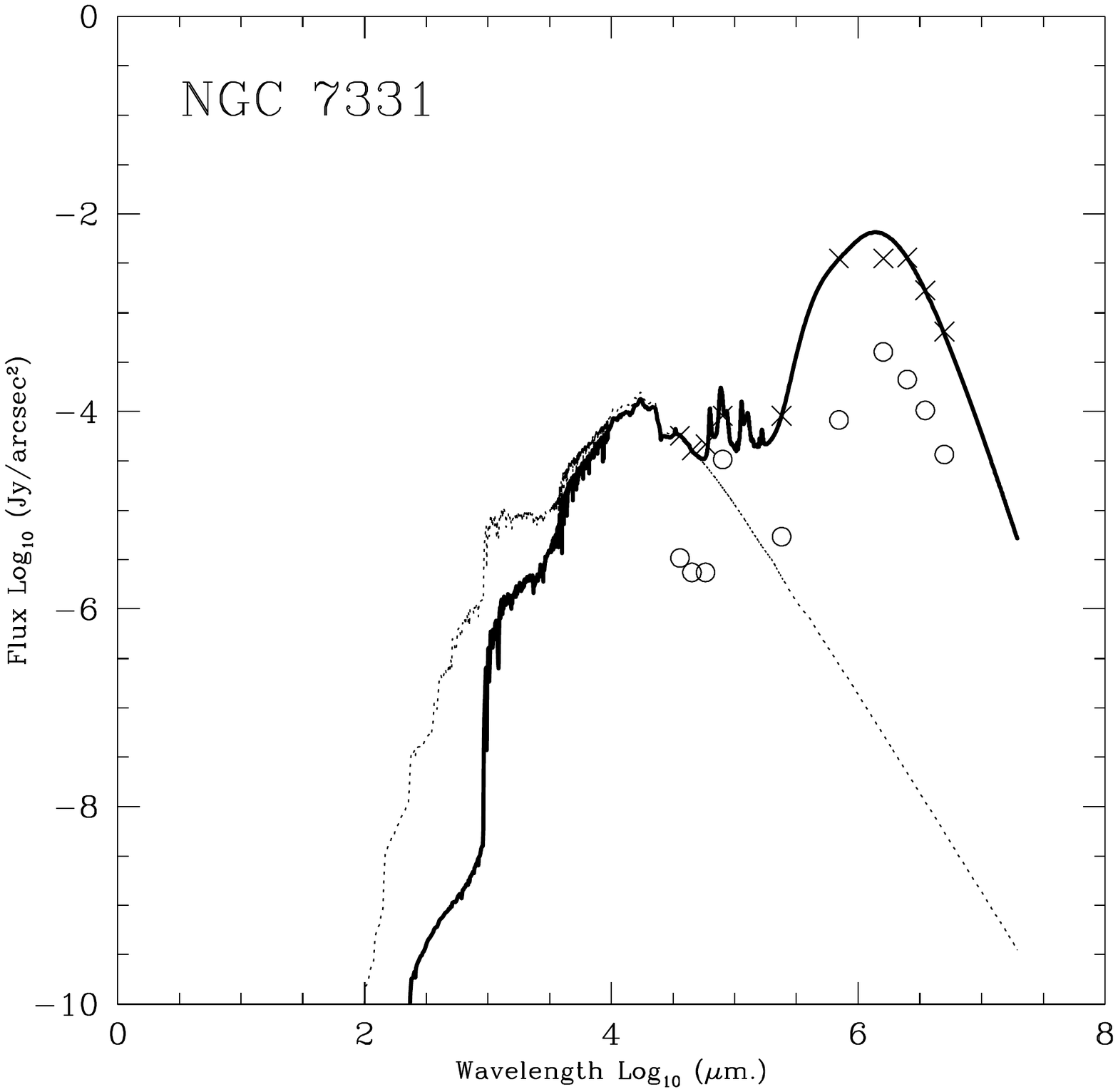}
\caption{The Spectral Energy Distribution from Spitzer IRAC and MIPS \citep{Holwerda07a} and the SPIRE from this paper. {\sc magphys} fits to this SED are plotted: the unattenuated spectrum (dotted) and the reprocessed spectrum (solid line). Open circles are the data-points, crosses the {\sc magphys} model values for each filter. }
\end{center}

\label{f:seds}
\end{figure*}

\end{document}